\documentclass[fleqn,usenatbib]{mnras}

\usepackage{newtxtext,newtxmath}
\usepackage[T1]{fontenc}
\usepackage{ae,aecompl}

\usepackage{graphicx}	% Including figure files
\usepackage{amsmath}	% Advanced maths commands
\usepackage{adjustbox}
\usepackage{xcolor}
\usepackage{titlesec}
\usepackage{subfigure}
\titleformat{\section}
       {\normalfont\fontfamily{phv}\fontsize{12}{15}\bfseries}{\thesection}{1em}{}

\titleformat{\subsection}
       {\normalfont\fontfamily{phv}\fontsize{10}{13}\bfseries}{\thesubsection}{1em}{}

\titleformat{\subsubsection}
       {\normalfont\fontfamily{phv}\fontsize{10}{13}\bfseries}{\thesubsubsection}{1em}{}

\title[The binary content of NGC 1850]{A closer look at the binary content of NGC 1850}%: a statistical analysis driven by individual systems}

\author[S. Saracino et al.]{S. Saracino$^{1}$\thanks{E-mail: s.saracino@ljmu.ac.uk},
S. Kamann$^{1}$, 
N. Bastian$^{2,3}$,
M. Gieles$^{4,5}$,
T. Shenar$^{6,7}$,
N. Reindl$^{8}$,\newauthor
J. M\"uller-Horn$^{9}$,
C. Usher$^{10}$,
S. Dreizler$^{9}$,
V. H\'{e}nault-Brunet$^{11}$\\
\\
$^{1}$Astrophysics Research Institute, Liverpool John Moores University, 146 Brownlow Hill, Liverpool L3 5RF, UK\\
$^{2}$ Donostia International Physics Center (DIPC), Paseo Manuel de Lardizabal, 4, 20018, Donostia-San Sebasti\'an, Guipuzkoa, Spain\\
$^{3}$ IKERBASQUE, Basque Foundation for Science, 48013, Bilbao, Spain \\
$^{4}$ ICREA, Pg. Llu\'{i}s Companys 23, E08010 Barcelona, Spain\\
$^{5}$ Institut de Ci\`{e}ncies del Cosmos (ICCUB), Universitat de Barcelona (IEEC-UB), Mart\'{i} i Franqu\`{e}s 1, E08028 Barcelona, Spain\\
$^{6}$Anton Pannekoek Institute for Astronomy, Science Park 904, 1098XH Amsterdam, The Netherlands\\
$^{7}$ Institute of Astronomy, KU Leuven, Celestijnenlaan 200D, 3001 Leuven, Belgium\\
$^{8}$ Institute for Physics and Astronomy, University of Potsdam, Karl-Liebknecht-Str. 24/25, 14476 Potsdam, Germany\\
$^{9}$ Institute for Astrophysics, Georg-August-University G\"ottingen, Friedrich-Hund-Platz 1, D-37077 G\"ottingen, Germany\\
$^{10}$ The Oskar Klein Centre, Department of Astronomy, Stockholm University, AlbaNova, SE-106 91 Stockholm, Sweden\\
$^{11}$ Department of Astronomy and Physics, Saint Mary’s University, 923 Robie Street, Halifax, NS B3H 3C3, Canada
}
\date{Accepted XXX. Received YYY; in original form ZZZ}
\pubyear{2023}

\begin{document}
\label{firstpage}
\pagerange{\pageref{firstpage}--\pageref{lastpage}}
\maketitle
\vspace{-0.5cm}
\begin{abstract}
Studies of young clusters have shown that a large fraction of O-/early B-type stars are in binary systems, where the binary fraction increases with mass. These massive stars are present in clusters of a few Myrs, but gradually disappear for older clusters. The lack of detailed studies of intermediate-age clusters has meant that almost no information is available on the multiplicity properties of stars with M$<$ 4$M_{\odot}$. In this study we present the first characterization of the binary content of NGC 1850, a 100 Myr-old massive star cluster in the Large Magellanic Cloud, relying on a VLT/MUSE multi-epoch spectroscopic campaign. By sampling stars down to M=2.5 $M_{\odot}$, we derive a close binary fraction of 24 $\pm$ 5 \% in NGC 1850, in good agreement with the multiplicity frequency predicted for stars of this mass range. We also find a trend with stellar mass (magnitude), with higher mass (brighter) stars having higher binary fractions. We modeled the radial velocity curves of individual binaries using \textsc{The Joker} and constrained the orbital properties of 27 systems, $\sim$17\% of all binaries with reliable radial velocities in NGC 1850. This study has brought to light a number of interesting objects, such as four binaries showing mass functions f(M)$>$1.25 $M_{\odot}$. One of these, star \symbol{35}47, has a peculiar spectrum, explainable with the presence of two disks in the system, around the visible star and the dark companion, which is a black hole candidate. These results confirm the importance and urgency of studying the binary content of clusters of any age.
\end{abstract}
%%%%%%%%%%%%%%%%%%%%%%%%%%%%%%%%%%%%%%%%%%%%%%%%%%

%%%%%%%%%%%%%%%%% BODY OF PAPER %%%%%%%%%%%%%%%%%%
\vspace{-0.5cm}
\section{Introduction}
\label{sec:intro}
Decades of observations have revealed that massive star clusters ($M>10^{4} M_{\odot}$) of all ages are key tools for studying the formation and evolution of stellar populations. Their detailed characterization has opened a window to previously unknown phenomena: from the existence of multiple stellar populations \citep[e.g.,][]{BastianLardo2018}, whose origin still lacks a coherent explanation, to the extended/bimodal main-sequence turn-offs (\citealt[e.g.,][]{Milone2015}, \citealt{Dantona2015}), a common feature of young clusters originating from different distributions of stellar rotational velocities (e.g. \citealt{BastianDeMink2009,Kamann2020,Kamann2023}).

Star clusters are very dense environments, in which interactions and collisions between stars tend to occur much more frequently than in the field. These interactions have a massive impact on the binary populations inside the clusters \citep{heggie1975}. While binaries with low binding energies tend to be quickly destroyed, tightly bound binaries are subject to a variety of phenomena, including fly-by or exchange interactions. This facilitates the formation of close binary systems and exotic objects, such as blue straggler stars, low-mass X-ray binaries, millisecond pulsars or cataclysmic variables \citep{Bailyn1992,Paresce1992,Ferraro1999,Ferraro2009} as well as peculiar rapidly rotating B-type stars like Be and shell stars \citep{pols1991,ZorecBriot1997,bodensteiner2020}. The multitude of interactions will cause the binary populations to strongly evolve with cluster age, resulting in evolving binary fractions, orbital parameter distributions, or pairing fractions. A clear indication of this process are the low binary fractions observed in ancient globular clusters \citep[e.g.][]{Milone2012}, relative to those found in the Galactic field for stars of the same mass \citep{MoeDiStefano2017}.

With the detection of gravitational waves from merging black holes (BHs) \citep{Abbott2016a,Abbott2016b}, the BH populations of star clusters have become a focus for the field. Given the high interaction rates and the tendency to form massive binaries via exchange interactions, star clusters appear as ideal environments for the dynamical formation of binary BHs \citep[e.g.][]{rodriguez2016,dicarlo2019}. In particular, merger cascades (or hierarchical mergers) appear feasible in star clusters, and offer a natural explanation for events such as GW190521 \citep{abbott2020}, involving BHs beyond the mass limits typically expected from stellar evolution. However, the merger rates inside star clusters are still uncertain \citep[see discussion in][]{mapelli2021}. Open questions regarding the multiplicity properties of massive stars or the retention of BHs following supernova kicks \citep[e.g.][]{atri2019} all contribute to this uncertainty.

Observationally, the direct evidence for the existence of BHs in star clusters is still limited, with a handful of objects in binaries with luminous companions having been reported thus far \citep{Maccarone2007b,Chomiuk2013,Strader2012,MillerJones2015,Giesers2018,Giesers2019}. In large part, this situation can be attributed to the challenges of performing spectroscopic studies in the crowded cluster environment\textbf{s}, limiting our abilities to detect BHs (or other types of remnants) in binaries with luminous companions. However, the advent of powerful integral field spectrographs like MUSE \citep{muse} is changing the landscape, as evidenced by the discovery of several BHs in the Galactic globular cluster NGC~3201 by \citet{Giesers2018,Giesers2019}.

\citet{Giesers2019} also highlighted the potential of MUSE to study the binary populations of star clusters as a whole and could constrain the period and eccentricity distributions of binaries inside NGC~3201 as well as the interplay between binarity and exotic objects such as blue stragglers and sub-subgiants. This raises the question if a similar approach can also be used to study young massive clusters, which may not only harbour a larger number of BHs compared to ancient globulars, but also show a number of peculiar features that have been linked to binarity. Examples in this respect are the unknown origin of the extended/bimodal main-sequence turn-offs \citep{Dantona2015,Bastian2020,Wang2022} or the rich populations of Be stars \citep{Bodensteiner2021,Kamann2023}.

Many spectroscopic studies have recently focused on characterizing the binary content of OB associations and very young open clusters, both Galactic and extra-galactic. These studies have mainly sampled O- and early B-type stars (massive stars, $\sim$10 $M_{\odot}$ or above) and have nicely shown that the binary fraction is the highest for the most massive stars, and then it rapidly decreases with decreasing mass. This is in agreement with the observational results of \citet{MoeDiStefano2017} and \citet{Sana2012}. For example, a bias-corrected close binary fraction of 51 $\pm$ 4\% and 58 $\pm$ 11\%, was found among young O- and B-type stars in the 30 Doradus region of the Large Magellanic Cloud (LMC, a few Myrs, \citealt{Sana2013,2015A&A...580A..93D}), of 52 $\pm$ 8\% among B-type stars in the young Galactic open cluster NGC 6231 (2-7 Myr, \citealt{Banyard2022}), of at least 40\% in the OB population of Westerlund 1 (10-20 Myr, \citealt{Ritchie2021}), of 34 $\pm$ 8\% among B-type stars in the SMC cluster NGC 330 (35-40 Myr, \citealt{Bodensteiner2021}) etc.
Over the years, further efforts have also been devoted to try and characterize the observational properties of individual binaries (in terms of period, eccentricity, mass ratio etc.). Interesting results in this direction were recently published, for example, for 30 Doradus \citep{Almeida2017,2021MNRAS.507.5348V} or Westerlund 1 \citep{2022A&A...660A..89R}. However, all binaries in these clusters are made of massive stars, of 10 $M_{\odot}$ or more. As a matter of fact, little or no information is currently available for binaries with intermediate-mass (2-5 $M_{\odot}$) stellar components. 

To fill this gap at intermediate masses, we initiated an observational campaign focused on analyzing the binary content of NGC 1850, a 100 Myr-old massive cluster in the LMC, using multi-epoch observations acquired with VLT/MUSE in two distinct observing runs. The target is ideal for two main reasons: 1) it is one of the most massive clusters in the LMC (M = 0.52 x $10^{\rm 5}$ M${_\odot}$, \citealt{song2021}), ensuring good statistics; 2) it belongs to a cluster age range, and corresponding star masses, yet to be explored in terms of their binary population. Furthermore, it is young enough so that little dynamical evolution has affected their binary properties, but yet old enough so that the population of exotic objects (e.g. BHs) should be already formed and  potentially present in the cluster. This dataset has already allowed to trace the rotational distribution and relative binary fraction of stars in the blue and red main-sequences (a common feature in clusters of this age, \citealt{Kamann2021,Kamann2023}) as well as to detect NGC1850 BH1, a binary system proposed to host a BH candidate \citep{Saracino2022}. This idea was later questioned \citep{2022MNRAS.511L..24E,2022MNRAS.511L..77S} but the nature of the unseen component is not yet established \citep{saracino2023}.

In this work we explore the binary content of NGC~1850 by sampling stars down to three magnitudes below the main-sequence turnoff of the cluster. Thanks to a statistically significant sample of binaries (of the order of hundreds), we can investigate the distribution of their main orbital properties (for example their periods). Observational studies as the one presented here are urgent and timely as they will provide important constraints for modelling massive star clusters, with the aim of obtaining more solid predictions in terms of their stellar and non-stellar population content. The urgency of such studies can be understood for example in the context of the controversy raised about the nature of NGC1850 BH1. In fact, \citet{2022MNRAS.511L..77S} used the binary evolution code BPASS to demonstrate that a binary like the one suggested by \citet{Saracino2022} should not exist in nature, being unsupported by stellar evolution. However, to reach this conclusion, the authors have used binary properties that are representative of the field, instead of massive clusters such as NGC 1850. In these systems, dynamical exchanges and interactions occur at a very high rate, with a strong impact on the evolution of the entire binary population and on the properties of individual binaries. How exactly they are influenced is poorly known.

The paper is structured as follows: in Section \ref{sec:obs} we briefly describe the observations and data reduction, we present the methods used to measure radial velocities and elaborate the approach used to identify binary systems. In Section \ref{sec:completeness} we present the sample and discuss the intrinsic binary fraction of NGC 1850, within the context of the MUSE field. Section \ref{sec:thejoker} is dedicated to the analysis of the radial velocity curves of binaries. The orbital properties and distributions of constrained binaries are presented in Section \ref{sec:binary_systems}. In Section \ref{sec:individual_systems} we describe the properties of individual binary systems, and elaborate on those with peculiar characteristics. After an extensive discussion of the results, we draw our conclusions in Section \ref{sec:concl}.
%\vspace{-0.5cm}
\section{Observations and data reduction}
\label{sec:obs}
A detailed description of the observations available for NGC 1850 can be found in previous papers using the same dataset (see \citealt{Kamann2021,Kamann2023} and \citealt{Saracino2022}). Here we limit ourselves to provide a brief description of the data, mainly focusing on their importance for detecting radial velocity variables (i.e. binary stars), which is the main purpose of this work.
The dataset of NGC 1850 consists of multi-epoch observations (Program IDs: 0102.D-0268 and 106.216T.001, PI: N. Bastian) acquired with MUSE \citep{muse}, the integral field spectrograph (IFS) mounted at the Very Large Telescope (VLT), in Chile. We used the wide field mode (WFM) configuration assisted by Adaptive Optics (AO), to improve the spatial resolution of the field and facilitate the extraction of single star spectra in such a high density environment. Since MUSE covers a field of view (FOV) of $1\times1$~arcmin at a spatial sampling of 0.2~arcsec in WFM, we used two pointings to sample the central regions of the cluster (the effective radius is indeed $r_{e}$ = 20.5 $\pm$ 1.4" (4.97 $\pm$ 0.35 pc), from \citealt{correnti2017}) and collect a statistically high number of stars. The two pointings overlap each other slightly, in order to gain deeper observations of that region (see Figure 1 in \citealt{Kamann2023}).

\begin{figure}
    \centering
	\includegraphics[width=0.45\textwidth]{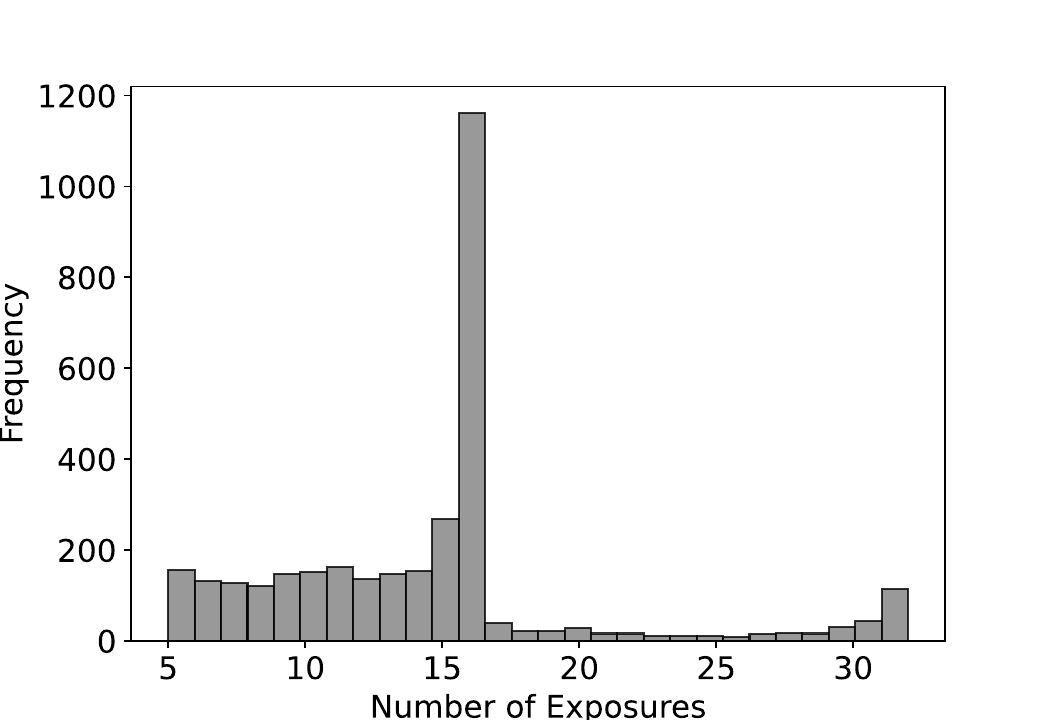}
    \caption{Histogram of the number of available MUSE observations per star. The vast majority of stars in NGC 1850 were observed 16 times (i.e. the number of MUSE epochs), meaning that one spectrum per epoch was extracted for each of them.}
    \label{fig:epochs}
\end{figure}

The observations cover a temporal baseline of more than two years (specifically 754.1 days), with a time sampling between individual epochs ranging from 1 hour to several months. We secured 16 observations per pointing and up to 32 independent observations for stars within the overlapping region. This strategy allows us to have an almost complete coverage of the inner regions of NGC 1850 both spatially and temporally, opening up the possibility of studying the kinematics of the stars within this massive star cluster. More quantitatively, the number of stars observed for a given number of epochs is presented in Figure \ref{fig:epochs}. The clear peak we see at 16 epochs corresponds to the default scenario where one spectrum per star/epoch is obtained. A small peak at 32 epochs is also observed and refers to the ``best observed" stars (i.e. stars in the overlapping region). A strategy that adopts non-uniformly sampled epochs allows to be sensitive to different types of binaries, and to discriminate between radial velocity signals of short and long period binaries. This is very important as it helps to avoid confusion caused by time aliasing and to be confident in determining the orbital periods of the binaries in the sample.
The reduction of the MUSE datacubes has been performed using the ESO MUSE pipeline \citep{pipeline} in the ESO Reflex framework. By default, the pipeline subtracts the sky lines and the sky continuum separately from the data. We did only perform the subtraction of the sky lines because the sky continuum in a crowded field like NGC 1850 ultimately contains a contribution of starlight that we did not want to remove. As a result, the final datacubes in our case still contain the telluric continuum emission and we model its contribution a posteriori, since it is spatially flat across the FOV and does not affect the spectra extraction.

Individual stellar spectra have been extracted from the MUSE FOV using the latest version of \textsc{PampelMuse} \citep{Kamann2013}, which relies on a Point Spread Function (PSF) fitting technique. We used a high-quality photometric catalog of NGC 1850 obtained from archival observations from the Hubble Space Telescope (HST) as a reference to determine both a MUSE PSF model and spatial coordinates of the resolved sources as a function of wavelength. This information is crucial to resolve and extract single sources even in the most crowded regions. We refer the reader to \citet{Kamann2023} for more details on this, including links to the publicly available photometric catalog and the combined MUSE spectra. 

Interestingly, NGC 1850 is one of the LMC clusters in the WAGGS (WiFeS Atlas of Galactic Globular cluster Spectra) project \citep{usher2017}. The WAGGS project exploits the Wide-Field Spectrograph (WiFeS) on the Australian National University (ANU) 2.3-metre telescope \citep{2007Ap&SS.310..255D,
2010Ap&SS.327..245D} which provides high resolution (R $\approx$ 6800) spatially resolved spectroscopy, with a wide wavelength range (3270–9050 \AA) covered by four gratings, of which two are observed simultaneously. The main aim of this project was to deduce the properties of star clusters through integrated spectroscopic studies however we exploited the capabilities of this instrument to obtain additional observations of NGC 1850 in order to measure the rotation of stars at its main-sequence turn-off as well as to look for radial velocity variability (PI: G. Da Costa). The observations acquired in September 2018 were specifically designed to have a perfect overlap to the MUSE FOV, in order to add further epochs to the radial velocity curves of bright targets (F438W $<$ 18.), as well as to have higher resolution spectra compared to MUSE. We extracted and analyzed the WiFeS data cubes of NGC~1850 with \textsc{PampelMuse} and the procedure described below applies consistently to both datasets to derive single-epoch radial velocities for each star in the FOV. We note here that the WiFes radial velocities will not be used to derive the binary fraction in NGC 1850. They will only be exploited as additional epochs to constrain the orbital properties of some binaries, where possible.
\vspace{-0.5cm}
\subsection{Radial velocities with Spexxy}
\label{sec:spexxy}
As described in \citet{Saracino2022} and \citet{Kamann2021,Kamann2023}, the extracted spectra were analysed with \textsc{Spexxy} \citep{Husser2016}, a python package which performs full-spectrum fits against a library of template spectra. The main aim of this process is to determine radial velocities and stellar parameters (e.g. effective temperature, metallicity) for every single source in the field. Rotational velocity measurements are not discussed in this paper as they have been already the subject of a dedicated study \citep{Kamann2023}. Given the presence in the field of a large number of hot stars with $T_{\rm eff}>$10,000 K, we used the synthetic templates from the \textsc{Ferre} library \citep{allendeprieto2018}, which includes spectra for stars with up to $T_{\rm eff}\sim$30 000 K. The initial guesses required by \textsc{Spexxy}, like surface gravity log~g, and effective temperature $T_{\rm eff}$ were taken from the comparison of the HST photometry to rotating and non-rotating MIST stellar models \citep{Gossage2019,mist2016}, assuming age of 100~Myr and metallicity of $[{\rm Fe/H}]$=-0.2, appropriate for NGC~1850. To avoid spurious measurements, we masked out any spectral ranges that could potentially be contaminated by the intense nebulosity that permeates the NGC~1850 field, including the cores of the strong Balmer lines (H$\alpha$ and H$\beta$).

We analyzed the \textsc{Spexxy} results a posteriori and considered as reliable all those that meet the following criteria: (1) The fit was marked as successful. (2) The derived signal-to-noise (S/N) of the input spectrum was $>5$. (3) The velocity determined by cross-correlating the spectrum with its best fit deviated by less than $3\sigma$ from the \textsc{Spexxy} result. (4) The \textit{Mag Accuracy} parameter returned by \textsc{PampelMuse}, which measures the agreement between the $m_{\rm F814W}$ magnitude derived from an extracted spectrum and the one in the HST photometry, was above 0.6. Spectra that did not meet these criteria were discarded, thus excluded from the final \textsc{Spexxy} catalog of MUSE radial velocities, which contains 3,283 stars.
\vspace{-0.5cm}
\subsection{Radial velocities with the cross-correlation}
\label{sec:cross}
In analyzing the radial velocities measured with \textsc{Spexxy}, we paid particular attention to two specific classes of objects: 1) stars for which \textsc{Spexxy} failed to provide a good full-spectrum fit, due to the presence of very peculiar spectral features, impossible to reproduce with any stellar template (see for example the case of star \symbol{35}47 and star \symbol{35}2413, discussed in Section \ref{sec:individual_systems}). As the analysis typically failed for such stars, they are not included in the \textsc{Spexxy} catalogue. 2) Stars that are heavily contaminated by nebular emission. For these stars, a full-spectrum fit could underestimate any radial velocity variations. A solution in such cases consists of analysing only the red part of the spectra`, i.e. the Paschen series, where nebula emission is negligible. For a detailed discussion of this issue, we refer to the case of star \symbol{35}224 or NGC1850 BH1 in \citet{saracino2023}.

To avoid unreliable or biased radial velocity measurements and also to recover the radial velocities of stars not analyzed by \textsc{Spexxy}, we decided to apply an alternative method to derive the relative radial velocities of the same NGC 1850 sample, which relies on cross-correlation (CC, hereafter) of the observations with a template spectrum created from the data themselves \citep{zucker1994,shenar2017,dsilva2020}. 
We performed the CC by exploiting the wavelength range between 7\,800~\AA\, and 9\,300~\AA. For each star, the spectrum with the highest S/N was used as an initial template. Afterwards, each remaining spectrum was cross correlated against the template. Then, a new template was created by shifting each spectrum by the inverse of its measured radial velocity and creating a S/N-weighted average of the shifted spectra. To avoid that individual low-S/N spectra impact the process, only spectra with $>40\%$ of the S/N of the initial template were included in the average. We repeated this process until the measured radial velocities converged. We note that this process results in relative rather than absolute velocities. When necessary, we converted to absolute velocities by analysing the final combined template with \textsc{Spexxy}. We applied the same S/N cut and the same magnitude accuracy as for \textsc{Spexxy}, to create the final CC catalog of MUSE radial velocities, which contains 2,662 stars. 

\vspace{0.2cm}
Since the two main goals of the study are: 1) to estimate the binary fraction in NGC 1850 and 2) to constrain the orbital properties of as many reliable binaries in the cluster as possible, we decided to select a ``high-quality'' sample of stars, for which we obtained completely consistent results with \textsc{Spexxy} and \textsc{CC}. All the results shown hereafter will be based on this final, clean hybrid catalogue.
A very efficient way to identify ``high-quality'' stars is to compare the binary probability $p$ derived from the two methods, instead of using individual radial velocities for each star. The binary probability, in fact, will be immediately correlated to how similar (or different) the results of \textsc{Spexxy} and \textsc{CC} are for that specific star. This selection will be discussed in the next Section, where the approach to estimate the binary probability for each star in NGC 1850 is presented.
\vspace{-0.5cm}
\subsection{Binary probability}
\label{sec:variability}
The main goal of this work is to investigate the presence of binary systems in the MUSE field of NGC 1850 and to characterize their properties, both in terms of overall distributions and of individual systems. For this reason, radial velocities determined from single-epoch spectra are the main information we are interested in. When looking for any signs of variability in the data, three different groups of stars can be identified: 1) stars in binary systems, where the variation in radial velocity is due to the motion of the star around a companion; 2) stars that are intrinsically variables, also called pulsators (RR Lyrae or Cepheids are common examples); 3) stars that show small variations due to the finite accuracy of our measurements but are not true variables. To perform this study we are interested in stars belonging to the first group, therefore an accurate knowledge of the radial velocity uncertainties is essential to discard all possible interlopers.

To calibrate the velocity uncertainties returned by both the \textsc{Spexxy} code and the CC method, we applied the approach presented in \citet{Kamann2020}. For each measured radial velocity in our sample we followed three steps: 1) first, we identified a sub-sample of 100 stars with similar $T_{\rm eff}$, log~g, and S/N values as the target star in the extraction under consideration (comparison sample). 2) Then, we measured the velocity differences of the stars in the comparison sample relative to every remaining epoch and divided the differences by the squared sums of the velocity uncertainties. Obvious binary stars were excluded from this comparison iteratively via k-sigma clipping. 3) Finally, we determined a correction factor for the formal uncertainty of the radial velocity measurement under investigation by measuring the standard deviation of the epoch-to-epoch distribution of the normalized velocity differences in the comparison sample. The fact that an independent correction factor can be obtained by comparing to each of the remaining epochs further allowed us to estimate the stability of the velocity calibration.

Velocities of evolved stars (F438W$<$18.0 and (F336W-F438W)$>$0.0) in NGC 1850 are measured to an accuracy between 1.0$\,{\rm km\,s^{-1}}$ and 2.0$\,{\rm km\,s^{-1}}$ during each epoch. On the main sequence, the typical uncertainties (per epoch) are larger, ranging from 5.0-6.0$\,{\rm km\,s^{-1}}$ at F438W $\sim$16. to 10.0$\,{\rm km\,s^{-1}}$ at F438W $\sim$18. to 25.0-30.0$\,{\rm km\,s^{-1}}$ for the faintest stars in our sample. Please note that the raw (i.e. before calibration) velocity uncertainties derived from the CC routine are on average larger (by a few percent, which is intrinsic to the method) than those obtained with \textsc{Spexxy}, so the calibrated uncertainties will also be larger in the latter case.

After carrying out an in-depth evaluation of the radial velocity uncertainties, the next step is to determine the probability that each star in the sample is a radial velocity variable. To do so we used the method developed by \citet{Giesers2019} for NGC 3201. For a given star observed for a number of epochs, we computed the reduced $\chi^2$ value for the set of radial velocity measurements and corresponding uncertainties under the null hypothesis of a single star (no radial velocity variations). The degrees of freedom are derived from the number of epochs available per star. This distribution over all observed stars is compared with what would be expected under the assumption that no radial velocity variables are available in the sample. More practically, for any given reduced $\chi^2$ value, this method computes the ratio between the number of stars \emph{observed} above this value and the number of stars \emph{expected} above this value in the absence of radial velocity variable stars. This comparison allows to assign a probability $p$ of being radial velocity variable to each star. High values of $p$ correspond to stars that are most likely part of a binary system, while low values of $p$ are generally associated to single stars. As a general assumption, we define as likely binary stars all those with $p>0.5$. This criterion will also be used later in the paper for creating the star catalogue that will be analyzed with the Bayesian software \textsc{The Joker}.

\begin{figure}
    \centering
	\includegraphics[width=0.475\textwidth]{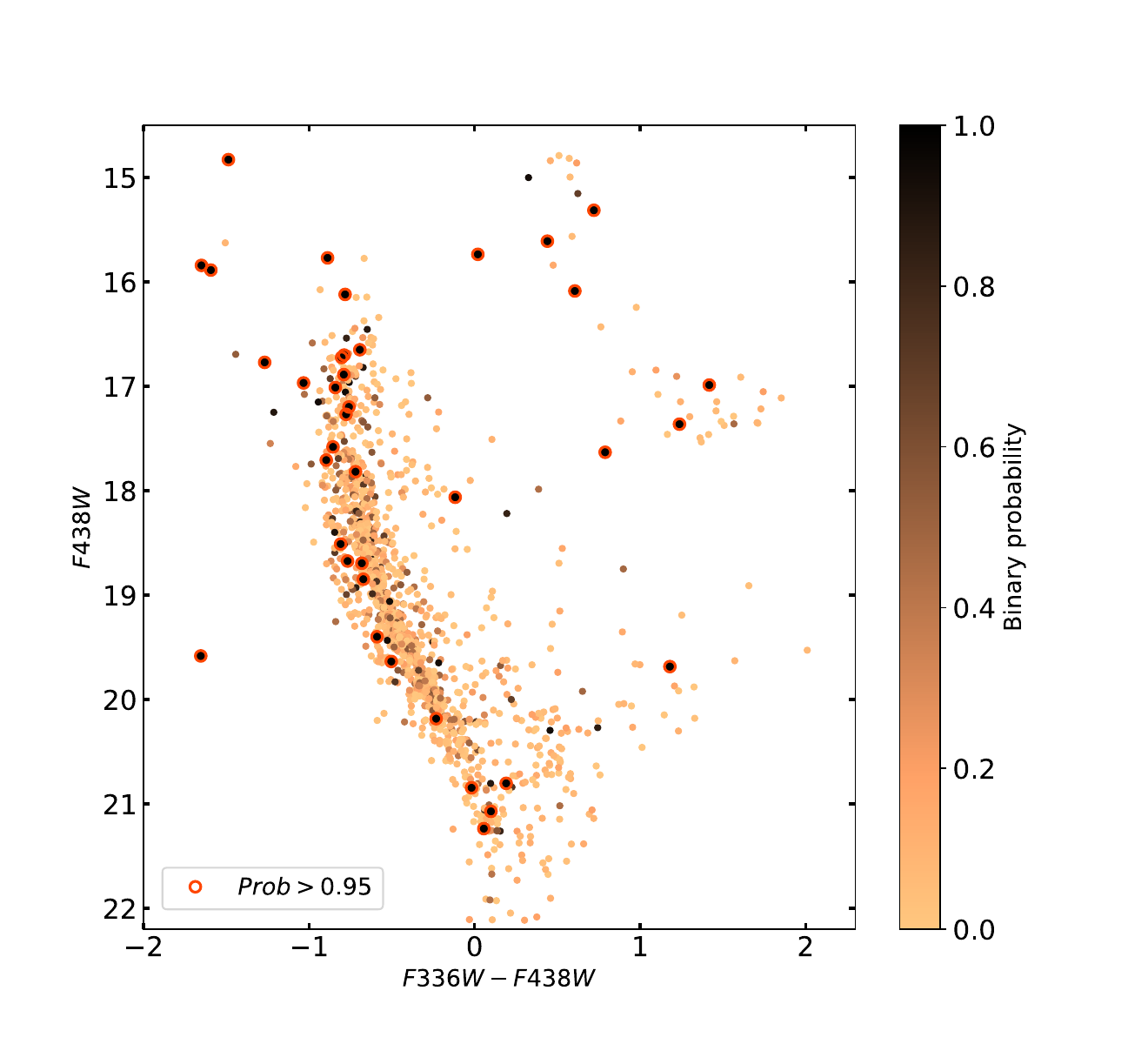}
    \caption{(F438W, F336W-F438W) CMD of NGC 1850. Each colored point represents a star with HST photometry, for which we have spectroscopic information from MUSE. The colors, from black to light orange refer to their probability of being radial velocity variables, thus being part of a binary system. Red open circles identify the stars in the sample with a probability p$>$95\% to orbit around a companion.}
    \label{fig:cmd}
\end{figure}

We applied this method to both the \textsc{Spexxy} and the CC results independently, in order to identify 1) which stars are considered binaries from both methods; and 2) which stars have ambiguous results, i.e. they are considered binaries by \textsc{Spexxy} but not by CC and vice versa. We realized that a non-negligible number of stars which were considered binaries (p$>$0.5) according to the \textsc{Spexxy} results, turned out to be single stars (p$<<$0.5) when processing the CC results. This is mainly caused by template mismatches (e.g. Be stars with emission lines), with the net effect of artificially inflating the fraction of binary stars in the inner regions of NGC 1850.
In order to get a clean sample of stars, with reliable radial velocities, we proceeded as follows:
%\vspace{-0.2cm}
\begin{itemize}
    \item First, we matched the stars in the two catalogs, finding 2,541 stars in common. To clean up this sample from stars with poorly measured radial velocities, we required the variability probabilities in the two approaches to be within 0.2 of each other. For example, if a star has p = 0.7 according to \textsc{Spexxy}, the same star analyzed with CC needs to have 0.5 $<$ p $<$ 0.9 in order to be included in the final, cleaned, catalogue. A sample of 1,441 stars met the criterion, corresponding to $\sim$57\% of the initial sample.
    \item Secondly, we looked at stars analysed by only one of the two approaches (\textsc{Spexxy} or CC). In particular, 742 stars from the \textsc{Spexxy} catalogue and 121 stars from CC. By looking at the magnitude distribution of these stars, we realized that stars in \textsc{Spexxy}-only sample are all very faint (F814W$>$20). These stars have low S/N spectra, which is the reason why they are not included in the CC catalogue. On the other hand, the CC-only stars cover the entire magnitude range, with an overabundance of stars at intermediate magnitudes (F814W$\sim$18-19). The latter are stars with peculiar spectra that cannot be fitted with standard stellar templates. 
    \item We created the final, clean catalogue, by adding to the 1,441 stars with consistent results between \textsc{Spexxy} and CC, the 121 stars from CC-only, obtaining a final sample of 1,562 stars. Although this sample does exclude a fraction of stars (mostly faint), it can still be considered a good representation of stars in the MUSE field of NGC 1850, i.e. those with reliable radial velocities over different methods. We will use \textsc{Spexxy} radial velocities for all stars measured by the two methods, because of the smaller uncertainties, while we will adopt radial velocities provided by the CC approach in all other cases. The CC technique is in fact the most reliable for stars belonging to the two classes mentioned above or to other peculiar categories. We have verified, however, that the results presented hereafter (in terms of binary fraction and orbital properties of constrained binaries) do not change if only the CC radial velocities are used.
\end{itemize} 
In Figure \ref{fig:cmd} we present the (F438W, F336W-F438W) colour-magnitude diagram (CMD) of NGC 1850, where each colored point is a star for which reliable MUSE radial velocities are available from the hybrid, clean catalogue. Stars most likely members of NGC 1850B are not shown in the figure. These stars were all located preferentially in the extension of the NGC 1850's main-sequence towards brighter magnitudes, overlapping the region where blue straggler stars are located. The binary probability distribution is shown as a colorbar, where the darker colors refer to single stars, while the lighter ones identify the binaries. Stars that are certainly binaries (with p$>$0.95) are also highlighted in the figure, with large red open circles.
\\
To enable the reader to more easily follow the remainder of the analysis, we provide below a brief summary of the stellar samples we will be using and the criteria used to generate them.
\begin{enumerate}
    \item[1)] To estimate the spectroscopic binary fraction of NGC 1850 we use:
    \begin{itemize}
        \item only cluster members (1,385 stars after removing 177 possible interlopers);
        \item stars with a minimum of 5 valid radial velocity measurements and p$>$0.5 are considered likely binaries;
        \item a magnitude cut (F814W$<$19.5) to ensure completeness of the photometric sample;     
    \end{itemize}
This selection results in 109 binary candidates, out of a sample of 1,033 stars remaining after the cuts. 
    \item[2)] To derive the properties of individual binary systems with \textsc{The Joker} we use:
    \begin{itemize}
        \item all stars in the MUSE FOV with reliable radial velocities (no membership cuts, 1,562 stars);
        \item stars with a minimum of 5 valid radial velocity measurements and p$>$0.5 are considered likely binaries;
        \item no magnitude cuts;      
    \end{itemize}
This selection results in 143 binary candidates, out of a sample of 1,562 stars remaining after the cuts.      
\end{enumerate}

\vspace{-0.5cm}
\section{The binary fraction of NGC 1850}
\label{sec:completeness}
Based on the statistical approach used to estimate the binary probability for each star in the sample, we can now compute the observed spectroscopic binary fraction for the inner regions of NGC 1850 in the MUSE FOV. 
The MUSE field is characterized by stars belonging to three different environments: 1) the massive, 100 Myr-old cluster NGC 1850; 2) the 5-Myr old, low mass, cluster NGC 1850B, responsible for the strong nebulosity observed in the field (see for example the right panel of Figure 1 in \citealt{Kamann2023}).
As this work focuses on the study of the binary fraction and the binary population of the main cluster NGC 1850, it is important to remove the contribution of the interlopers, in order to have a sample of cluster members as clean as possible. NGC 1850 stars and stars belonging to the field share similar radial velocity values, thus a distinction based on radial velocities remains ambiguous. In fact, \citet{song2021} has measured for NGC 1850 a systemic velocity of $v_{\rm sys}$ = 248.9 $\,{\rm km\,s^{-1}}$ and a velocity dispersion of $\sigma$ = 2.5 $\,{\rm km\,s^{-1}}$, while for the LMC field a systemic velocity of $v_{\rm field}$ = 257.4 $\,{\rm km\,s^{-1}}$ with a dispersion of $v_{\rm field}$ = 23.6 $\,{\rm km\,s^{-1}}$. The latter is also consistent with the uncertainties provided by other literature works studying the LMC field, for example with the GAIA data \citep{vasiliev2018}.

Likely members of NGC 1850B and the LMC field can actually be identified on the basis of photometric and/or spatial properties. In fact, from previous works in the literature, as well as from a visual inspection of the cluster CMD in specific filters, the contamination from field stars appears not severe. In addition, LMC stars are generally much older ($\sim$6 Gyr) than NGC 1850 stars, so the LMC main-sequence overlaps with NGC 1850 only at rather faint magnitudes, starting at about F438W$\sim$21.5, corresponding to the faintest stars in the MUSE sample, which are not included in the final clean catalogue. To be more specific, from Figure \ref{fig:cmd}, field stars are predominantly located at F438W$>$18. and colors (F336W-F438W)$>$0.2-0.25.
The photometric argument does not really work for NGC 1850B, since its main-sequence partially overlaps with that of the main cluster, NGC 1850. However, its contribution in stars can still be inferred by looking at the (RA-Dec) parameter space: these two clusters are very close but still distinguishable in the sky. All stars within 10" of a visually determined center ($\alpha=05^{\rm h}08^{\rm m}39.3^{\rm s}$, $\delta=-68^\circ 45^{\prime} 45\farcs5$) can be considered likely members of NGC 1850B. 

To determine the observed spectroscopic binary fraction of the main cluster NGC 1850 we applied the photometric and spatial cuts mentioned above to remove the most probable members of NGC 1850B and the LMC field from the sample of 1,562 stars with reliable radial velocities. This is safe to do, since this result will not be significantly affected by a few mis-classified stars. However, we anticipate here that we will run \textsc{The Joker} on the entire dataset, as we want to be as conservative as possible when looking for the orbital solutions of each individual binary in the cluster.

\begin{figure}
    \centering
    \includegraphics[width=0.48\textwidth]{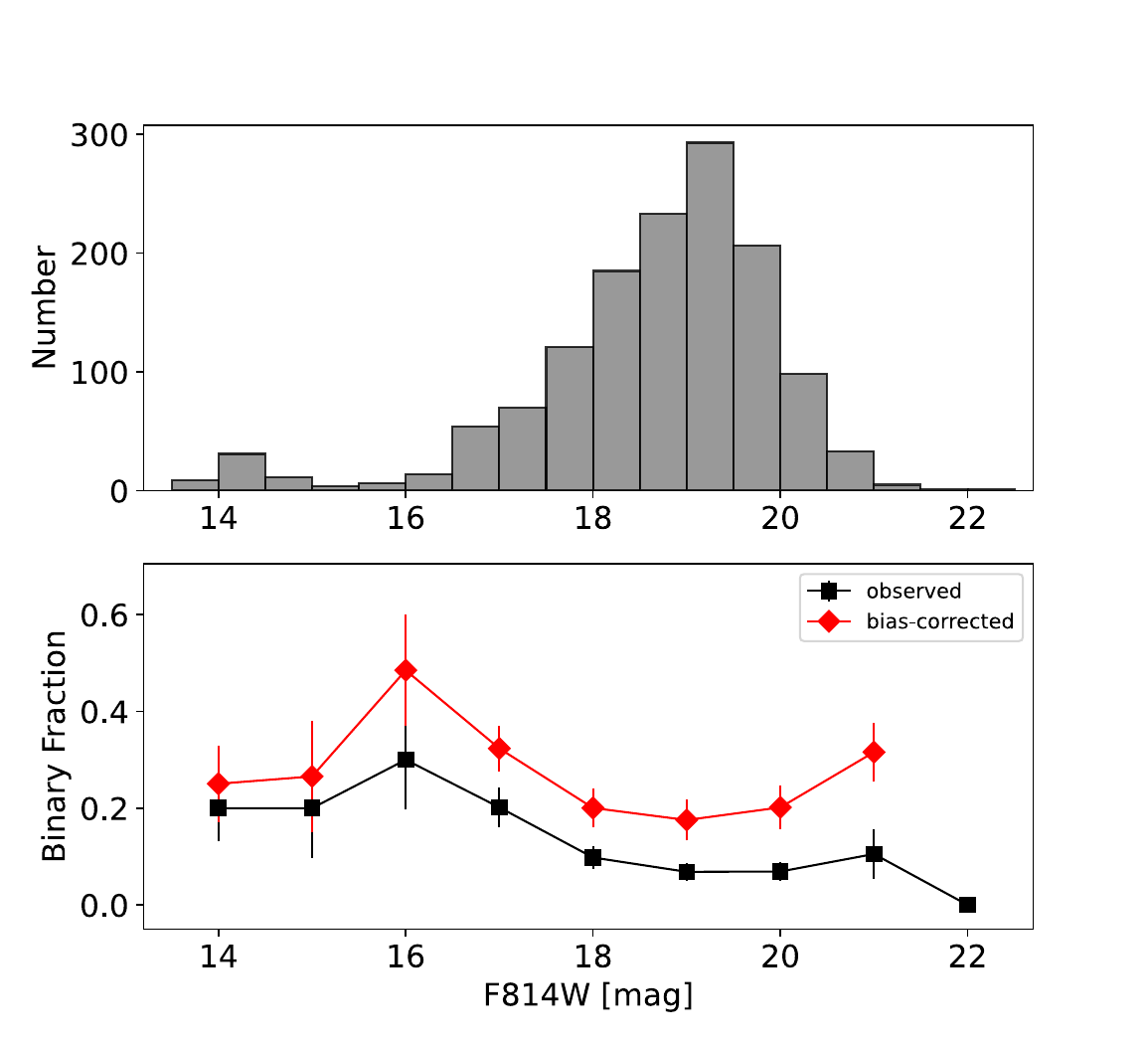}
    \caption{Distribution of F814W magnitudes and observed spectroscopic binary fraction for stars in NGC 1850. {\it Top panel:} distribution of F814W magnitudes of all stars with reliable radial velocity measurements (dark gray). {\it Bottom panel}: observed spectroscopic binary fraction as a function of F814W magnitude (black points). All stars with binary probability $p>=0.5$ are considered binaries here. See text for more details. By excluding the lower and higher magnitude bins, where low number statistics and incompleteness respectively play an important role, a trend is clearly observed, i.e. the binary fraction increases to brighter magnitudes. After correcting for the detection probability, the intrinsic binary fractions for different magnitudes are also plotted as red diamonds. Interestingly, the same trend with magnitude is observed.}
    \label{fig:binaryfraction_mag}
\end{figure}

\begin{figure*}
    \centering
    \includegraphics[width=0.85\textwidth]{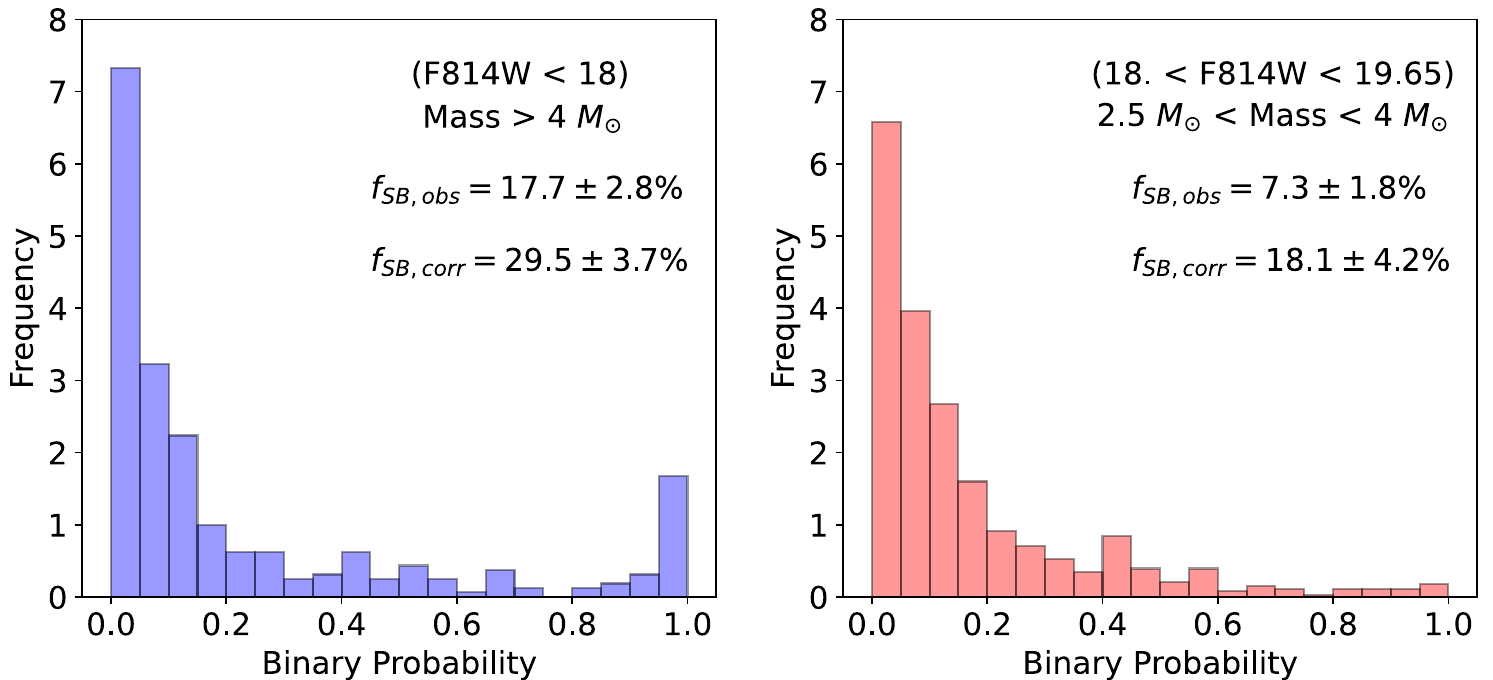}
    \caption{Histograms of the binary probability for stars in two different mass regimes in the NGC 1850 clean sample: {\it Left panel:} Stars with high masses (M$>$4 $M_{\odot}$, corresponding to F814W magnitudes $<$ 18.). {\it Right panel:} Stars with intermediate masses (2.5 $<$ M $<$ 4 $M_{\odot}$, corresponding to 18. $<$ F814W $<$ 19.65 mag). As indicated in the plots, the binary fraction is very different in the two regimes, with the high-mass end featuring significantly more binaries. This is also observed when the bias-corrected binary fractions are measured (also labeled in the plots).}
    \label{fig:binaryfraction_mass}
\end{figure*}

The catalogue of members of NGC 1850 consists of 1,385 stars with reliable radial velocity measurements. All stars are detected at least in 5 MUSE epochs. Figure \ref{fig:binaryfraction_mag} shows the F814W magnitude distribution of all stars in this sample, as well as the corresponding binary fraction, by adopting bins of 1 mag each. The observed binary fractions are obtained from the comparison between the number of stars with $p>0.5$ in each magnitude bin and the total number of stars with reliable radial velocity measurements in that bin. The statistical uncertainties of the binary fractions are calculated using the prescriptions of \citet{Giesers2019}: by the quadratic propagation of the uncertainty determined by bootstrapping (random sampling with replacement) the sample and the difference of the fraction for $p>0.45$ and $p>0.55$ divided by 2 as a proxy for the discriminability uncertainty between binary and single stars.
In the upper panel of Figure \ref{fig:binaryfraction_mag} two important aspects need to be pointed out: a) the bins corresponding to the brightest magnitudes contain a limited number of stars, leading to larger uncertainties in the binary fractions due to limited number statistics; 2) for magnitudes F814W$>$19.5, the number of stars per bin rapidly decreases. This is where our spectroscopic sample starts to become incomplete. The binary fraction increases for the faintest stars: binaries are brighter, so they will be over-represented in this incomplete sample. If we focus on the binary fraction for stars at magnitudes 15.5 $<$ F814W $<$ 19.5, we then observe a decreasing trend, with the binary fraction varying from 30\% to $\sim$7\%.
An alternative way to visualize this difference is by plotting the binary probability distribution for stars in two different mass regimes: stars with high masses (M $>$ 4 $M_{\odot}$\footnote{A star in NGC 1850 can have a maximum mass of about 5 $M_{\odot}$. This corresponds to the upper mass limit in this regime.}, corresponding to F814W $<$ 18.) in the left panel of Figure \ref{fig:binaryfraction_mass} and stars with intermediate-masses (2.5 $<$ M $<$ 4 $M_{\odot}$, corresponding to 18. $<$ F814W $<$ 19.65 mag) in the right panel of the same Figure. As already suggested by the observed trend in magnitude of Figure \ref{fig:binaryfraction_mag}, the binary fraction in the high-mass end is 17.7 $\pm$ 2.8 \%, more than double the one in the intermediate-mass regime, 7.3 $\pm$ 1.8 \%.

Finally, to determine the overall spectroscopic binary fraction of NGC 1850 in the MUSE FOV, we take into account all stars with F814W $<$ 19.5, to assure that our spectroscopic sample is complete. We derive an observed binary fraction of 10.6 $\pm$ 1.8 \%, meaning that in a sample of 1,033 stars, we find 109 stars with $p>0.5$, i.e. likely binary systems. We also investigated how the observed spectroscopic binary fraction varies as a function of the distance from the center of NGC 1850, since our dataset samples nearly three effective radii of the cluster ($\sim$60"). Although the uncertainties are significant, we observe that the binary fraction decreases with increasing distance from the center, approximately from 12\% for the innermost bin to 9\% for the outermost bin. This is consistent with the mass segregation phenomenon already observed in other clusters (e.g. \citealt{Giesers2019}). We note here that variable stars (e.g. Cepheids or RR Lyrae) are partially inflating these numbers, as there is no way to distinguish them at this level from true binaries. The sample will be cleaned {\it a posteriori} of variable stars (mainly Cepheids) thanks to cross-matching with archival catalogs (e.g. OGLE-IV). A discussion on this aspect can be found in Section \ref{pulsators}, but the low number of detected variables in the MUSE FOV does not have any impact on the overall binary fraction of NGC 1850.
\vspace{-0.5cm}
\subsection{The binary fraction among Be stars}
\label{sec:Bestars}
In very young clusters stars are used to be classified as not-evolved, i.e. still on the main-sequence, and evolved, i.e. giant stars that have already evolved off the main-sequence. In NGC 1850, the latter group is significantly under-populated compared to the former, so providing the binary fractions for the two groups separately is not very informative. However, unlike other clusters of similar age, NGC 1850 is known to host a significant number of Be star candidates (see \citealt{milone2018}, \citealt{correnti2017}), i.e. rapidly rotating B-type stars with a circumstellar decretion disk that gives rise to strong Balmer-line emissions \citep{Rivinius2013}. Exploiting the same dataset we are using here, \citet{Kamann2023} have studied this population of stars in detail from a spectroscopic perspective, by selecting the sample of Be stars with reliable radial velocity measurements but also discriminating between ``classical Be stars" and shell stars, i.e. Be stars observed through their disks (viewed at very high inclination angles), thanks to a detailed inspection of their spectra. In this work we add another piece of information to the Be star picture, by studying the observed spectroscopic binary fraction of Be star candidates in NGC 1850 and compare it with the fractions found in other clusters in the literature.

\begin{figure}
    \centering
	\includegraphics[width=0.49\textwidth]{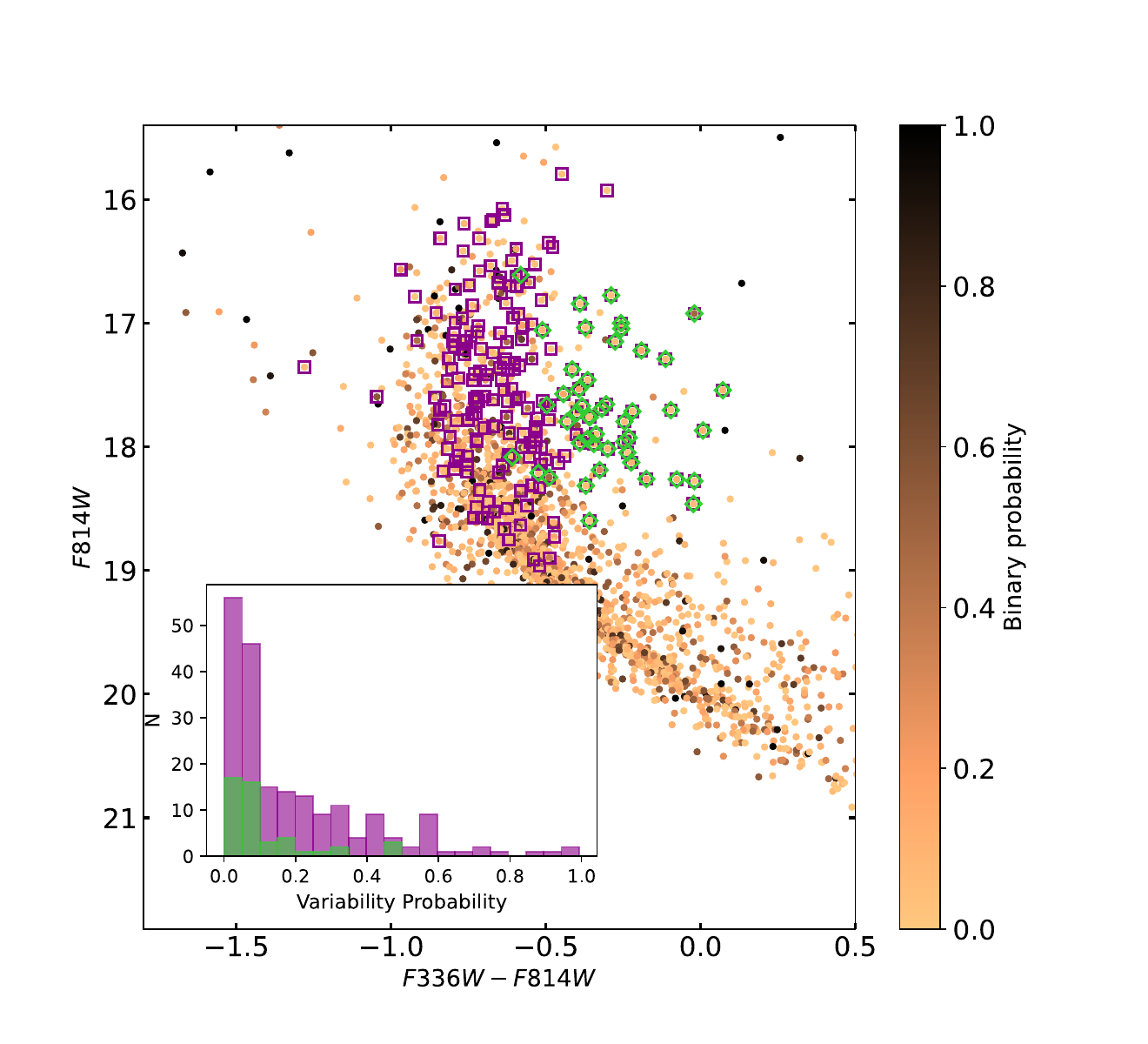}
    \caption{(F814W, F336W-F814W) CMD of NGC 1850. As in Figure \ref{fig:cmd}, each colored point represents a star with HST photometry, for which we have spectroscopic information from MUSE. The colors, from black to light orange refer to their probability of being radial velocity variables, thus being part of a binary system. Purple open squares and green diamonds identify Be and shell star candidates, respectively, in NGC 1850 according to the spectroscopic investigation made in \citet{Kamann2023}. The binary probability distributions of Be and shell star candidates are shown in the inset (purple and green, respectively). The fraction of obvious binaries (i.e. p>0.8) in the two groups of stars is quite low or equal to 0.}
    \label{fig:bestars}
\end{figure}

Figure \ref{fig:bestars} shows the CMD of NGC 1850 in a specific combination of HST filters (F814W, F336W-F814W) that allow us to easily identify Be stars in the cluster and clearly separate shell stars. The spectroscopic sample of Be and shell star candidates from \citet{Kamann2023} are highlighted by large red squares (202 stars) and large green diamonds (47 stars), respectively, while the color code for all stars is the same as in Figure \ref{fig:cmd}.
The inset in Figure \ref{fig:bestars} shows the binary probability distribution of all Be star candidates in red and of the sub-sample of shell stars in green. We find 20 Be stars with $p>0.5$, measuring an observed spectroscopic binary fraction of 10.0 $\pm$ 2.8 \%, while we do not find any shell star showing signs of variability ($f_{\rm obs,shell}$ = 0\%). In order to understand how these fractions compare with the fraction of binaries for other classes of objects in NGC 1850, we did not apply a completeness correction factor\footnote{Many Be stars are known to be post-interaction products, so the prescriptions we adopt in Section \ref{sec:comp_corr} to estimate the completeness factor for all stars in NGC 1850 would probably not be realistic here.}, but we performed a relative comparison to stars in the main-sequence of the cluster, at the same magnitude range of Be stars. The amount of completeness is expected to be the same for stars in the same magnitude range. For this sub-sample of main-sequence stars we derived an observed spectroscopic binary fraction of 11.0 $\pm$ 1.9 \%. Within the uncertainties, Be and main-sequence stars have similar binary fractions in NGC 1850. 
This is interesting as it slightly differs from what was found in other clusters. In NGC 330, for example, \citet{Bodensteiner2021} found that the observed binary fraction of Be stars was nearly half that of main-sequence stars (15.9\% vs 7.5\%). It is worth noticing, however, that if we only focus on obvious binary systems ($p>0.8$ or higher), we do find $f_{\rm obs, MS}$ exceeding $f_{\rm obs, Be}$. On the other end, the very low binary fraction they found among Be stars $\sim$0.3 mag redder that the main-sequence (2\%, see their Fig. 7), i.e. what we call shell stars, is in excellent agreement with what we find here. It is not entirely clear yet why shell stars are mostly observed as single stars, however there might be a few effects at play. \citet{Bodensteiner2021} for example suggested two possibilities: a) the close binary fraction of classical Be stars is intrinsically low; or b) Be binaries have long orbital periods, for which the detection probability drops. Only a more detailed investigation of this class of objects can help shed light on this aspect.
\vspace{-0.5cm}
\subsection{Correction for observational biases}
\label{sec:comp_corr}
In order to compare the overall binary fraction of NGC 1850 with other studies in the literature, but also to investigate the reliability of the magnitude/mass trends observed (i.e. whether they are caused by our ability of detecting binaries at different magnitude levels, or they are linked to an intrinsic difference in the multiplicity properties of NGC 1850's stars at different magnitudes/masses), a proper understanding of the different biases affecting our measurements is needed. The main limitations of our observational campaign are the following: 1) Not all orbital periods have the same probability of being detected. In particular, due to the temporal sampling of our survey, we are more sensitive to tight binaries than to long period ones; 2) Many binaries have orbital velocities below our detection threshold, so they may have been classified as single stars when they are not. 3) The method we use is actually biased against detecting binaries of (almost) equally bright stars. In such cases, both binary components contribute (almost) equally to the observed spectrum. In the combined spectrum, the radial velocities of the two components mostly cancel out, making it extremely challenging to detect such binary systems based on low-resolution spectroscopy. In a recent work with MUSE data on the 35-40 Myr-old cluster NGC 330, \citet{Bodensteiner2021} found that their probability to detect binaries with a mass ratio q$>$0.8 drops to 10\% or less, due to the phenomenon just described.

To test our ability in detecting binary systems given our observational setup (e.g. in terms of time coverage, number of epochs, radial velocity uncertainties), we performed two different sets of simulations. The main difference between the two is on the distribution of the parent populations adopted for three orbital parameters, such as period, mass ratio (q=$M_{2}/M_{1}$) and eccentricity. We adopted the same distributions as in previous works of this kind (e.g. \citealt{Bodensteiner2021} and references therein) to allow a consistent comparison of the results. For the first set we assumed a log-uniform period distribution ranging from 0.15 to 3.5 (i.e., P from 1.4 to 3160 days), an eccentricity distribution between 0 and 0.9 proportional to $\rm \sqrt{e}$, with a circularization correction for periods $<$ 2 days (see discussion in Section \ref{binary_population} and Eq. 1) and a flat mass ratio distribution between 0 and 1 (see the finding in \citealt{2022A&A...665A.148S} for O-type stars). In the second set of simulations we instead adopted power law distributions for all three parameters: $\rm (logP)^\pi$ with $\pi$ = -0.25 $\pm$ 0.25 for the period \citep{Bodensteiner2021}; $\rm q^k$ with $k$ = -0.2 $\pm$ 0.6 for the mass ratio and $\rm e^\eta$ with $\eta$ = -0.4 $\pm$ 0.2 for the eccentricity \citep{Sana2012}. The exponents were drawn from normal distributions with central values and 1$\sigma$ dispersions taken as described above.

We generated 1,385 binary systems (the number is chosen to be equal to the number of stars in the cleaned sample of NGC 1850), by randomly assigning orbital parameters (e.g. period, eccentricity, mass ratio etc.) to each of them from the parent populations presented above. We further adopted a random orientation of the orbital plane in 3D space and a random reference time $t_{0}$. Of all binary systems thus created, we discarded those with unphysical solutions based on two criteria: 1) {\it binary hardness}, i.e. how much the system is bound. The equation adopted in the simulations is Eq. 5 from \citet{ivanova2005}, which determines whether the binary system survives the cluster environment based on its binding energy. All systems with binary hardness $<$ 1 are treated as single stars. 2) {\it Roche Lobe overflow}, i.e. when the radius of the most massive star in a binary exceeds its Roche limit. When this happens, the system is in the common envelope phase and can no longer be treated as a binary in this context.
In order to take into account the effect on the radial velocities of SB2 binaries, or binaries with almost (equally) bright components, we also implemented the damping factor formula derived by \citet{Giesers2019} (see their Eq. 1), as they found that the theoretical expected radial velocities are linearly damped with the flux ratio of the two components. The closer the mass ratio q is to 1, the closer the combined radial velocities are to the systemic velocity, completely erasing the signal and making these types of systems very difficult to detect.

Once all radial velocities were generated for all binaries surviving the two criteria ($\sim$ 94\% of the initial sample), we adopted the same approach described in Section \ref{sec:variability} to estimate the binary probability of each system in the sample, then its binary fraction, comparing the systems with $p>0.5$ to the total number of surviving binaries. Since these are simulations, we know exactly how many binaries and single stars we have in the sample and this allows us to estimate the detection probability, i.e. how good we are in detecting binaries given the properties of our observations. For each of the two configurations we have iterated the simulations 10,000 times, to allow a robust calculation of the detection probability, but also to have a good idea of the associated uncertainties.

We obtained a detection probability of 43.4 $\pm$ 4.7 \% from the analysis of the first set of simulations, while $\rm 45.1^{\rm -15.3}_{\rm +9.2}$ \% for the second set. These values were obtained by computing the 50th percentile of the distributions generated by the 10,000 simulations. When using these detection probabilities to correct the observed spectroscopic binary fractions, we obtain a true binary fraction for stars in NGC 1850 of 24.4 $\pm$ 5.0 \% and $\rm 23.5^{\rm -15.4}_{\rm +9.4}$ \% for the two sets of simulations, respectively.
The uncertainties associated with the detection probability and, as a consequence, to the true binary fraction of NGC 1850 are made of two terms: 1) a systematic uncertainty due to the number of simulations performed; 2) a statistic uncertainty due to the binary population sample size. We obtained the former by calculating the 16th and the 84th percentiles ($\pm$ 1$\sigma$) of the above distributions\footnote{We do not use the mean and standard deviations of the distributions, as they are not perfectly gaussian, but have a longer tail in one direction.}. To estimate the latter, we instead ran two more sets of simulations (of 10,000 realizations each), this time using the true binary fractions as input parameters to the simulations. The number of binaries in these simulations corresponds to the intrinsic number of binaries in the NGC 1850 sample, thus the width of the error distributions (always $\pm$ 1$\sigma$) reflects the impact of the sample size.

Although the two sets of simulations have very different parent population distributions, the intrinsic, bias-corrected, binary fractions we obtain for NGC 1850 are remarkably similar to each other, considering their uncertainties. The latter are significant for the simulations with power law distributions for period, mass ratio and eccentricity but this is not surprising given the large variety of exponents that could be randomly assigned to these parameters. Indeed, they could switch from distributions in period that favor tight binaries ($\pi$ $\sim$ -0.5), which are easy to detect with our observational setup, to those that favor much longer orbital periods ($\pi$ $\sim$ 0), very difficult to detect. This has the net effect of increasing the spread in the detection probabilities, therefore increasing the uncertainties. The shape of the period distribution is by far the dominant factor that affects the resulting detection probabilities, as already noted by \citet{Bodensteiner2021}. 

\begin{figure}
    \centering
    \includegraphics[width=0.48\textwidth]{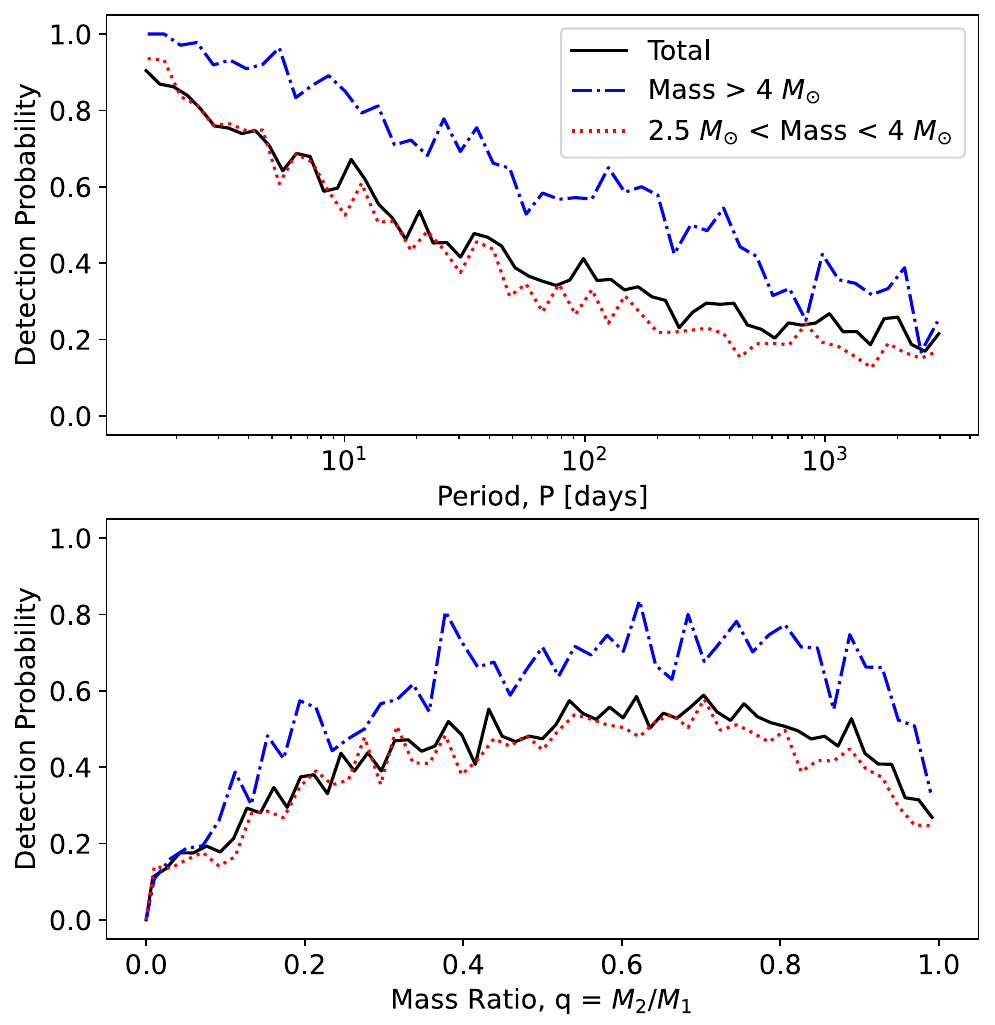}
    \caption{Binary detection probability simulations based on the observational setup of our survey. {\it Top panel:} Detection probability as a function of the orbital period P (computed assuming a flat mass-ratio distribution) for the total number of stars (in black) and for two different mass regimes: stars with primary mass $M_{1}>$ 4 $M_{\odot}$ in blue and stars with intermediate masses (2.5 $< M_{1} <$ 4 $M_{\odot}$) in red. {\it Bottom panel:} Detection probability as a function of mass ratio q (computed assuming a flat $\rm log P$ distribution) shown as a black curve. The blue and red curves define the two mass regimes mentioned above. The simulations take into account all possible biases that reduce or alter our sensitivity in detecting binaries. Binaries with long orbital periods are the most difficult to detect, as are binaries of near-equal mass. More details can be found in the text. The binary detection probability is significantly higher for high-mass stars, due to their brightness. The uncertainties on the radial velocity measurements are small for such stars, as their spectra are among those with the highest S/N. This increases our efficiency in detecting radial velocity variations.}
    \label{fig:parameters}
\end{figure}

To visually inspect the results of our simulations, in Figure \ref{fig:parameters} we show how the detection probability varies as a function of the orbital period and mass ratio of the binaries. We consider only the first set of simulations (with flat $\rm logP$ and mass ratio distributions), and use 10 realizations. We include all binaries in the sample, without applying any cut. We note here that by increasing the number of simulations used, the results would not change, only smoother curves would be obtained. 
As can be seen, the detection probability decreases with increasing orbital period as our observational setup makes them more difficult to detect. It ranges from 0.9 for periods of a few days to 0.2 for periods of thousands of days. 
For the mass ratio, a very low detection probability ($\sim$ 0.1) is observed for binaries with low q values, reaching a peak around 0.5 for q=0.5-0.7, then decreasing for $q>0.8$. The observed behaviour is consistent with what found by \citet{Bodensteiner2021} in their Fig. 2 (orange lines) using a similar set of simulations. In this work, however, we observe that the drop in the detection probability is significantly less drastic than in the simulations by \citet{Bodensteiner2021}. It is reasonable to link this discrepancy to differences in the observational setup between the NGC 1850 dataset and the one available for NGC 330.

When we do the same exercise as in Figure \ref{fig:parameters} but we now consider two different mass regimes, i.e. high-masses: M $>$4 $M_{\odot}$ and intermediate-masses: 2.5 $<$ M $<$ 4 $M_{\odot}$, we obtain the results presented respectively as a blue and a red curve in the same figure. Overall, the detection probability is higher in the high-mass regime (blue), than in the intermediate-mass regime (red). This is expected since more massive stars are also brighter in the CMD; they have high S/N spectra, and their radial velocities are measured with lower uncertainties, allowing binaries with significantly lower amplitudes to be detected than in the other regime. 
To investigate whether the trends observed in Figure \ref{fig:binaryfraction_mag} as a function of magnitude and in Figure \ref{fig:binaryfraction_mass} as a function of mass are still present after correcting for the observational biases just mentioned, we calculated the true binary fractions to compare with the observed spectroscopic binary fractions. The bias-corrected values are shown as red diamonds in Figure \ref{fig:binaryfraction_mag} and are labeled in Figure \ref{fig:binaryfraction_mass}. Although the differences between the different regimes have become less important, trends in magnitude/mass can still be observed, which may actually reflect an intrinsic difference in the multiplicity properties of these stars.
\vspace{-0.5cm}
\subsection{Comparison to other clusters}
The close binary fraction in the 100 Myr-old cluster NGC 1850 can be directly compared to the fraction measured in other clusters in the literature. As mentioned in Section \ref{sec:intro}, most of the clusters analyzed so far are actually very young (of the order of a few or few tens of Myrs) and still contain a large fraction of O- and early B-type stars with masses up to $\sim$10 $M_{\odot}$ or more, see for example 30 Doradus, NGC 6231, Westerlund 1 etc. NGC 1850 is the oldest massive cluster for which a detailed spectroscopic analysis of the binary content has been performed. Due to the massive stars involved, the close binary fraction in these clusters is much higher (over 50\% in 30 Doradus \citep{Sana2013,2015A&A...580A..93D} and NGC 6231 \citep{Banyard2022} and over 40 \% in Westerlund 1 \citep{Ritchie2021}) than what we derived in this work for NGC 1850, but if we also include NGC 1850 in the sample we clearly observe a trend with mass.
A much closer comparison in terms of true binary fraction can actually be made with the 35-40 Myr-old cluster NGC 330 in the low metallicity environment of the SMC, Z$\sim$1/5 $Z_{\odot}$ \citep{Bodensteiner2021}. This cluster has a binary fraction of 34 $\pm$ 8\% but the observations sample stars down to 4 $M_{\odot}$. Although the overall binary fraction in NGC 1850 is still lower than what the authors find in NGC 330, if we compare the binary fraction of NGC 1850 in the high-mass regime (left panel of Figure \ref{fig:binaryfraction_mass}), where only stars with masses of 4 $M_{\odot}$ and above are included (thus setting the same low-mass threshold in the two clusters), the two close binary fractions come into agreement (34 $\pm$ 8\% vs 29.5 $\pm$ 3.7\%). This result confirms that the binary fraction in a cluster is indeed mass-driven, with other cluster properties such as metallicity or age potentially influencing these values as well, but much less strongly.
\vspace{-0.5cm}
\section{Determination of binary properties}
\label{sec:thejoker}
The statistical approach applied in the previous sections has provided us with a sample of binary candidates, i.e. all stars with binary probability $p>0.5$. For each of these objects we have an observed radial velocity curve which, although only sparsely sampled, contains much information about the orbital properties of the binary itself. 
Several methods can be used to infer these properties, and one of the most used in recent years is the Generalised Lomb-Scargle periodogram (GLS; \citealt{Zechmeister2009}). It is very effective for finding the orbital period of binary systems but usually works best when uniformly sampled datasets are available. Considering the nature of our data, this approach would only be successfully applied to a small subset of stars, particularly those with a populated radial velocity curve, where the orbital period is fairly well sampled. In all other cases, unfortunately, the GLS method would not be ideal.

For that reason, we exploit a slightly different approach here, which was only recently developed and has already proven to be very successful with similar sets of data. Called \textsc{The Joker}, this tool is a custom Monte Carlo sampler for sparse or noisy radial velocity measurements of two-body systems, and can produce posterior samples for orbital parameters even when the likelihood function is poorly behaved \citep{Price-Whelan2017,Price-Whelan2020}. We applied this software to our NGC 1850 dataset with the goal of constraining the properties of individual systems, whose sampling and number of epochs are good enough to unequivocally determine their orbits. 
\vspace{-0.5cm}
\subsection{The Joker}
Literature studies have demonstrated that at least 5 epochs are needed to constrain the orbit of a binary system \citep{Price-Whelan2017}. Therefore, in order to run \textsc{The Joker}, we focused only on stars with at least 5 reliable MUSE spectra, corresponding to 5 radial velocity measurements, and, as mentioned previously, with a binary probability $p>0.5$. We used the entire dataset, without any {\it a priori} selection based on the probability of the stars to belong to the main cluster NGC 1850.

To make \textsc{The Joker} work properly on our dataset, we also made two main assumptions: 1) all binaries are made of two stars, so no triple or multiple systems are included in our modeling. 2) in our binary systems, one of the two stars always dominates in light with respect to the other. This means that we have modeled all binaries as being SB1, even though we know this assumption is not entirely correct. SB2 stars, i.e. stars in which the two components are equally bright and contribute almost equally to the system's light, in fact, do exist in clusters like NGC 1850 but can be identified only after a proper inspection of the stellar spectra and in general with higher spectral resolution. 

The idea behind \textsc{The Joker} is to create a huge library of possible orbits, based on a set of input parameters, such as period, eccentricity, semi-amplitude velocity K etc. When the radial velocity curve of a star is provided, the software scans the defined parameter space to find the family of orbits that best match the observed radial velocity curve. If this procedure is successful and the orbit is well constrained by the data, only one set of possible orbits (all with very similar orbital periods) is shown, providing a unique solution. Otherwise, depending on the behaviour of the data, \textsc{The Joker} can provide a few possible solutions or a set of solutions randomly distributed over the entire allowed period range. The latter means that the orbit of that specific binary is poorly constrained or not at all constrained, respectively.

In particular we generated $2^{29}$ = 536 870 912 prior samples, uniformly distributed within a period range between 1~d and 1,000~d. We did not include any orbital solution shorter than 1~d as this would have introduced an artificial bias toward unphysical, very short period binaries. We checked whether some of the stars in our sample showed a preference for periods below the lower limit we imposed but couldn't find any within the constrained sample (see Section \ref{sec:individual_systems} for details). For the upper limit we assumed a period slightly larger than our temporal baseline, in order to give enough room to the software to find the appropriate solution. Binaries with orbital periods longer than 1,000 d may indeed be present in the cluster but would probably have such a low semi-amplitude velocity K that it would be impossible for us to detect them as binaries.

As for the eccentricity and the velocity semi-amplitude K, we have followed the prescriptions given in Table 1 of \citet{Price-Whelan2020}, i.e. a beta distribution for e and a normal distribution centered in 0 $\,{\rm km\,s^{-1}}$ and with $\sigma$ = 30 $\,{\rm km\,s^{-1}}$ for K. However, we have carried out several tests by modifying the distribution for some of these parameters (e.g. log-normal distribution vs uniform distribution in P or standard vs custom normal distribution in K) and we get very consistent results for the constrained binaries, which gives us confidence that the results do not depend on the type of distributions adopted.

We assumed a normal distribution for the systemic velocity of the binaries, centered on $v_0=250.0\,{\rm km\,s^{-1}}$, which is close to the systemic velocity of NGC~1850 and with $\sigma=20.0\,{\rm km\,s^{-1}}$, which is consistent with the velocity dispersion of LMC field stars \citep{vasiliev2018}. The systemic velocity of the binaries in NGC 1850 is expected to follow the distribution with the dispersion of the cluster ($\sigma=5.0\,{\rm km\,s^{-1}}$), however, this choice is conservative because it ensures that we cover all cluster binaries but do not bias our prior against velocity of field binaries, thus avoiding the misinterpretation of some results. On the other hand, it gives us the possibility to identify {\it a posteriori} those binaries that are likely part of the cluster, by assigning them a membership probability through photometric and/or spatial location in the MUSE FOV.

We requested 512 posterior samples (solutions) per star and, as mentioned above, we analyzed only stars with a minimum of 5 epochs and a binary probability $p>0.5$ or higher. As soon as one star gets fewer than 256 posterior samples (half of the total), two different strategies are applied, depending on how constrained the solution is. 1) If the solution is nearly unimodal, we use a dedicated Monte Carlo Markov chain (MCMC) run as described in \citet{Price-Whelan2017,Price-Whelan2020} to get 512 new samples. 2) If the solution is more widespread, we perform an iterative procedure, generating new samples until reaching the minimum number of 256. The results of the entire process, both in terms of the binary population properties and of individual peculiar systems are presented in the following sections.
\vspace{-0.5cm}
\section{The binary population of NGC 1850}
\label{sec:binary_systems}
We ran \textsc{The Joker} on a sample of 143 binary systems fulfilling the criteria described in Section \ref{sec:thejoker}. This analysis can result in three very different outcomes: 1) a unimodal distribution, i.e. the solution is unique and the orbital parameters are well determined; 2) a bimodal distribution, i.e. two sets of solutions are possible, thus the orbit is close to be constrained; 3) a multi-modal or continuous distribution, i.e. the orbital parameters are poorly constrained or completely unconstrained. Among those, we only focused on objects with unimodal or bimodal posterior distributions. 

In order to discriminate between unimodal, bimodal or unconstrained solutions, we applied the method described in \citet{Giesers2019}. For each star we computed the standard deviation of the posterior periods $P$ returned by \textsc{The Joker} on a logarithmic scale. Stars with $\sigma_{\ln{P}}$ $<$ 0.5 are identified as unimodal and represent the most constrained systems in our sample. We find that 20 out of 143 binaries have unimodal solutions. An example of a binary with a unimodal solution is presented in Appendix \ref{app:extra} (Figure \ref{fig:unimodal}) for star \symbol{35}260, where the phase-folded radial velocity curve\footnote{To compute the phase-folded radial velocity curve in this plot and the plots hereafter we took the median in period and returned the values of the other parameters for that sample, as done in \textsc{The Joker}.} is also presented, to show how good the agreement between the data and the best-fit model is. Bimodal posterior samples were instead determined by using a k-means clustering algorithm with k=2 from the \textsc{scikit-learn} python package \citep{Pedregosa2011} to separate two sets in the period posterior distribution, where each of the two sets has to fulfill the criterion above. The set which includes the largest number of samples is used as the preferred solution in these particular cases.
To determine the best-fit orbital period per binary, we take the 50th percentile of its distribution as the median and the 16th and 84th percentiles as 1$\sigma$ respectively below and above the median. Starting from this period range ($P_{\rm min}$ - $P_{\rm max}$), we then extract the corresponding distribution in all other parameters (e.g. eccentricity, semi-amplitude velocity, systemic velocity etc.). The best-fit values for each of these parameters are finally determined as the 16th, 50th and 84th percentiles of the newly created distribution. We find bimodal solutions for 1 out of 143 binaries in our sample. The binary with a bimodal orbital solution is star \symbol{35}303 and is presented in Figure \ref{fig:bimodal}. 

The results of \textsc{The Joker} in terms of star position (RA and Dec coordinates), $m_{F438W}$ magnitude and best-fit values of the fitted parameters (period, eccentricity, semi-amplitude K, systemic velocity $v_{\rm sys}$, etc.) are listed in Appendix \ref{app:extra} in Table \ref{tab:properties} for all binaries with constrained solutions. The last column helps identifying which binary belongs to which group (unimodal vs bimodal). To verify the accuracy of the constrained solutions derived by \textsc{The Joker}, we analyzed the data using two more approaches. In particular we applied the GLS code mentioned above, as well as the UltraNest software \citep{buchner2021} and we obtained for all stars listed in Table \ref{tab:properties} consistent results within their uncertainties. 

All the remaining 116 binaries in the MUSE sample have multi-modal or continuous distributions (their properties are not constrained), hence they cannot be used to describe the properties of the binary population in NGC 1850.
\vspace{-0.5cm}
\subsection{Binaries constrained with MUSE + WiFeS data}
In the sample of binaries analyzed with \textsc{The Joker}, 6 out of 143 deserve particular attention. These binaries are all characterized by orbital periods which are preferentially longer than 100 d. Although such a period is well within the temporal baseline of our MUSE observations, the strategy we have used to design the MUSE campaign is such that the data is much more sensitive to binaries with orbital periods of the order of a few days or tens of days with respect to hundreds. For this reason, the solutions provided by \textsc{The Joker} for these systems are all multi-modal when considering only MUSE epochs. Adding new epochs to the sample is really the only way to extend the temporal baseline of our dataset and more importantly its sampling, in order to be able to constrain even the longest orbits of these systems. By exploiting the complementary WiFeS program of NGC 1850 (PI: G. Da Costa), we were able to extract the spectra of some bright stars in the WiFeS FOV. More specifically, the brightness of our targets (F438W $<$ 18.) combined with the perfect overlap between the WiFeS data and the MUSE FOV made it possible to add further epochs to the radial velocity curves of 64 binary stars. In this way, much better constrained orbital properties were derived for each of the 6 binaries mentioned above.
The remaining 58 binaries can be easily divided into two categories: a) binaries whose orbital properties were already well constrained using the MUSE data; and b) binaries that remained unconstrained even after the WiFeS data were added.
\vspace{-0.3cm}
\subsection{Orbital properties of constrained binaries}
\label{binary_population}
Almost 17\% of the total number of binaries detected in the MUSE FOV of NGC 1850 have well constrained orbital properties, in terms of orbital period, eccentricity, semi-amplitude velocity K etc. These properties are listed, along with their uncertainties, in Table \ref{tab:properties} and include all 27 binaries with unimodal and bimodal solutions, both with MUSE and MUSE+WiFeS. The number of binaries with constrained orbits was actually larger, but the final sample was created by removing photometric variable stars, e.g. pulsators, which we identified by cross-matching our dataset with the OGLE-IV catalog (a discussion on this topic is presented in the next sections).

We tried to assign each of the 27 constrained binaries a probability of being a member of NGC 1850, NGC 1850B or the LMC field, and found that 25 out of 27 are very likely members of the massive cluster NGC 1850. Indeed, based on the HST photometry, there are no stars with color (F336W-F438W)$>$0.5 and magnitude $m_{F438W}$>18., so we can safely conclude that none of them are part of the LMC field. However, there are two stars (\symbol{35}32 and \symbol{35}51) that are spatially within a radius of less than 10" from the center of NGC 1850B. This leads us to believe that these stars are probably members of NGC 1850B, although this is not certain, as many members of NGC 1850 may still occupy that region.

\begin{figure}
    \centering
	\includegraphics[width=0.48\textwidth]{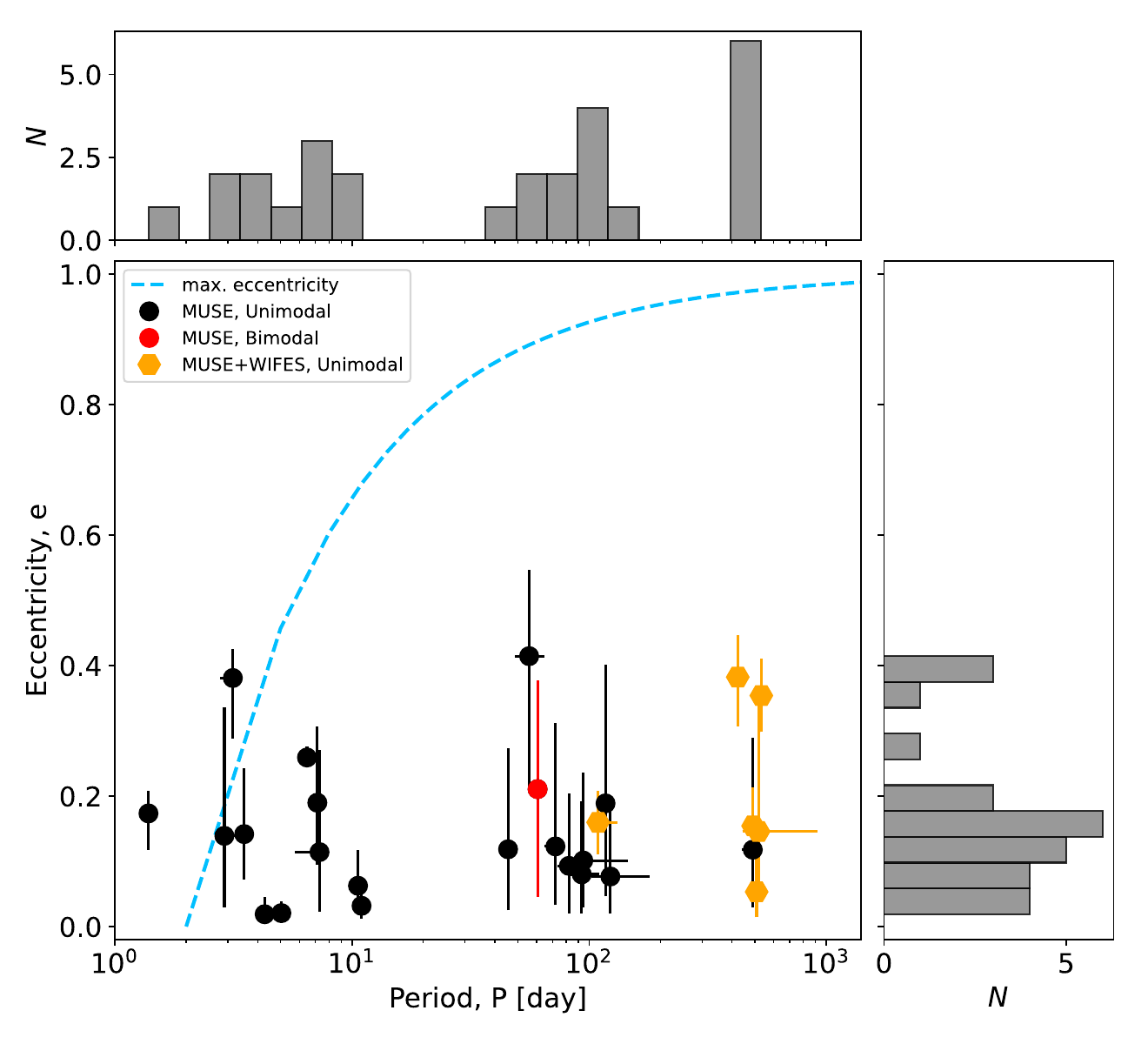}
    \caption{Eccentricity - Period plot of the well constrained binaries in NGC 1850. Binaries with unimodal and bimodal solutions in the posterior period sampling are shown as black and red dots, respectively. Those are derived using only MUSE measurements. Stars with unimodal solutions obtained with MUSE+WiFeS data are instead presented as orange pentagons. The period distribution of the 27 binaries is shown in gray in logarithmic scale and spans the range between 1 and 500 days with multiple peaks. The eccentricity distribution, on the other hand, varies only from 0 to 0.4, with a peak around 0.1/0.15, i.e. prefers low eccentricity orbits. The dashed cyan line defines the parameter space in which binaries with eccentric orbits should be located. Binaries with P$<$ 2 days are expected to have circular or close to circular orbits. All constrained binaries in NGC 1850 appear to follow this trend nicely, except one.}
    \label{fig:distribution}
\end{figure}

The eccentricity - period distribution of the 27 constrained binaries in NGC 1850 is presented in a linear-log plot in the main panel of Figure \ref{fig:distribution}. Black dots identify binaries with unimodal posterior samples, while the red dot is the binary with bimodal posterior samples, as obtained by using the MUSE epochs. Orange hexagons instead indicate binaries with unimodal solutions obtained thanks to the addition of further epochs from the WiFeS dataset. Eccentricity and period distributions are also shown as gray histograms in the vertical and horizontal panels, to better visualize the results.

We find binaries in a rather large period range, from $\sim$1.3 d to $\sim$500 d, but their period distribution does not uniformly cover the entire range. It shows, instead, multiple peaks, corresponding to the period intervals in which our observations, due to the temporal sampling of the MUSE campaign, were more sensitive. The peak at $\sim$ 400-500 d is almost entirely produced by the binaries constrained thanks to the WiFeS data. Overall, the $\pm$1$\sigma$ uncertainty in period is rather small, except for a few binaries which have significantly larger uncertainties (though we note that the x-axis is logaritmic). The latter, despite being classified as unimodal or bimodal based on the criterion presented at the beginning of the section, show posterior samples clustered in more than one group at periods very close to each other. This affects the width of the distribution, thus the uncertainties. 

The eccentricity distribution of the constrained binaries, on the other hand, covers only a narrow range of values from 0 to 0.4, with a broad distribution and perhaps the hint of a peak around 0.1-0.15. This suggests that the binaries in our sample have relatively low eccentricities. However, eccentricity is the least constrained orbital property of all. Consequently, with the exception of a very few outliers, any value of e between 0 and 0.4 is allowed for most binaries. Finally, there are no highly eccentric binaries (e$>$0.5) in the sample. It is not surprising to see that the eccentricity distribution is biased towards low eccentricities, because for such orbits fewer radial velocity measurements are necessary to get a unique solution. 

\begin{figure}
    \centering
	\includegraphics[width=0.48\textwidth]{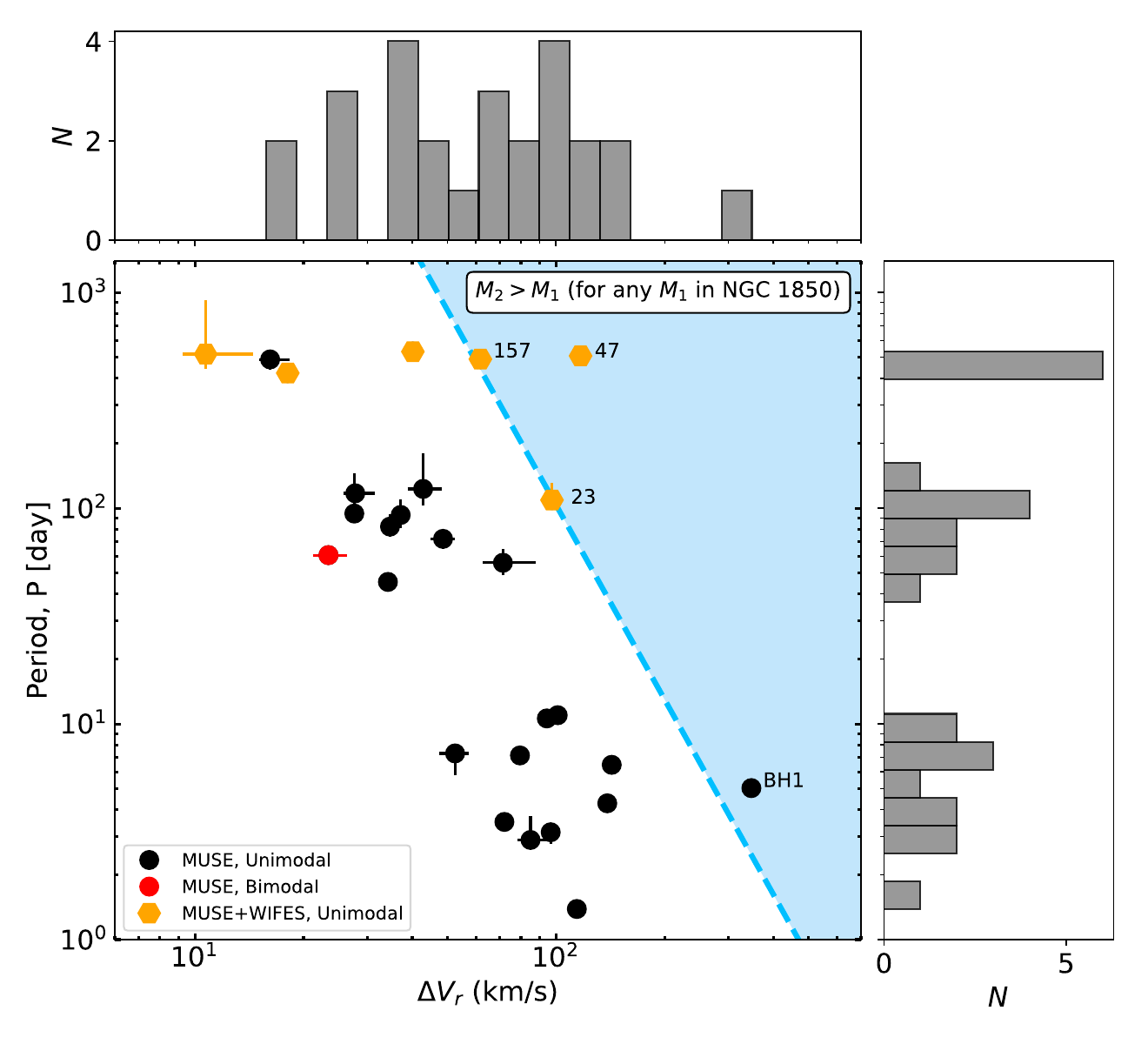}
    \caption{Period - Peak-to-Peak radial velocity variation ($\Delta V_{r}$) plot of the 27 well constrained binaries in NGC 1850. The colour code is the same as in Figure \ref{fig:distribution}, also shown in the bottom-left legend. Stars with large orbital periods and/or high peak-to-peak radial velocity variability can be easily identified. The cyan shaded area is the region in which stars with massive companions are expected to be located. In particular, the cyan dashed line identifies the locus where equal-mass binaries (in particular $M_{1}=M_{2}=5M_{\odot}$) are located, i.e. this is the largest mass that a star in NGC 1850 can have (specifically, at the turn-off), given its absolute age. A few constrained binaries fall in this area, one of which is the previously discussed BH candidate NGC1850 BH1 \citep{Saracino2022,saracino2023}. They are labeled individually in the figure, being the subject of further discussions later in the text. Finally, the 1D period and $\Delta V_{r}$ distributions of the 27 binaries are also shown in the figure, in gray in logarithmic scale, spanning a large range of values.}
    \label{fig:distributionK}
\end{figure}

In Figure \ref{fig:distribution} we also show as a cyan dashed line a maximum eccentricity $e_{\rm max}$ power law derived from a Maxwellian thermal eccentricity distribution:
\begin{equation}
    e_{\rm max}(P) = 1 - \left(\frac{P}{2\,{\rm days}}\right)^{2/3}\;\;\;\; {\rm  for}\; {\rm P\,}>2\, {\rm days}
\end{equation}
for a given period P \citep{MoeDiStefano2017}. This represents the binary components having Roche lobe fill factors $\leq70$\% at periastron. They predict that all binaries with P $<$ 2 d should have circular orbits due to tidal forces. It is interesting to note that, with the exception of one star (star \symbol{35}1828, see Table \ref{tab:properties}), all the others behave accordingly.
The period distribution presented in Figure \ref{fig:distribution} is now plotted against the peak-to-peak radial velocity distribution ($\Delta V_{r} = 2K$) of the 27 binaries, in a log-log plot in Figure \ref{fig:distributionK}. The colors are the same as in Figure \ref{fig:distribution}. Here we observe a large range of amplitude values among the binaries, specifically from tens of $\,{\rm km\,s^{-1}}$ to hundreds of $\,{\rm km\,s^{-1}}$, with one system even exceeding 300 $\,{\rm km\,s^{-1}}$. The properties of this peculiar binary system NGC1850 BH1 have been already discussed (see \citealt{Saracino2022,saracino2023,2022MNRAS.511L..24E,2022MNRAS.511L..77S}).

According to \citet{Clavel2021}, this plot can be used to identify the region where stars with massive companions are possibly located. A main-sequence turn-off star in NGC 1850 is of $\sim$5 $M_{\odot}$, which is the maximum mass a star in such a cluster can assume, according to its absolute age (100 Myr). We used this assumption to derive the position, in the (P-$\Delta V_{r}$) parameter space, of a binary made up of two components: 1) a primary star as massive as a 5 $M_{\odot}$ star (the star we observe), and 2) a secondary component with exactly the same mass, so that the mass ratio of the system (q=$M_{2}/M_{1}$) is equal to 1. The locus of points with such a property is identified by the cyan dashed line in Figure \ref{fig:distributionK}. 

All binary systems having q=1 but masses for the individual components lower than 5 $M_{\odot}$ define lines parallel to the left of the cyan dashed line (the lower the mass, the more they move to the left). Conversely, the cyan shaded region to the right of the cyan line is populated by binary systems in which the secondary (or invisible) component is significantly more massive than the primary (or observed) star (for $M_{1}\leq5M_{\odot}$, which is an appropriate assumption for stars in NGC 1850). Binaries with massive invisible companions are very interesting as they are the best candidates for hosting dark compact objects, such as NSs and BHs. In our sample of 27 constrained binaries, there are four that fall in this region, with two being very close to the dashed line. As labeled in the plot, these binaries have HST ID \symbol{35}23, \symbol{35}47, \symbol{35}157 and \symbol{35}224. We label the latter as BH1 instead, to adopt the same nomenclature as the papers that have discussed its properties \citep{Saracino2022,saracino2023}. We will discuss the properties of star \symbol{35}23, star \symbol{35}47 and star \symbol{35}157 in detail in a dedicated section (\ref{sec:individual_systems}).
The binary mass function, f, is an important orbital property we can derive directly once we know the period, the eccentricity and the semi-amplitude velocity K of a binary system, thanks to the formula:
\begin{equation}
f = \frac{P\,K_{\rm 1}^{\rm 3}(1-e^{2})^{3/2}}{2\pi G},   
\end{equation}
where G is Newton’s gravitational constant.
Without knowing {\it a priori} the mass of the visible source, this quantity can instruct us on the properties (e.g. mass) of the source that we do not detect. In fact, using Kepler's third law, the binary mass function f(M) can also be written in the form:
\begin{equation}
f(M_{1},M_{2}) = \frac{M_{\rm 2}^{3}\sin(i)^3}{(M_{\rm 2}+M_{\rm 1})^2},   
\label{eq:fm2}
\end{equation}
where $M_{\rm 1}$ and $M_{\rm 2}$ are the masses of the primary visible star and the secondary unseen component, respectively, and $i$ the inclination angle of the binary with respect to the line of sight. In massive star clusters, where we know the maximum mass a star can assume, the binary mass function f(M) is a very informative parameter since, without making any assumptions on the mass of the observed star and/or the inclination of the system, it is able to predict which of the two components should be more massive. In NGC 1850, for example, where the maximum mass for a star is $\sim$ 5 $M_{\odot}$, if we consider the most conservative scenario, i.e. the binary system is nearly edge-on ($i$ $\approx$ 90$^\circ$), we obtain that the secondary (unseen) component is surely more massive than the primary (visible) component if f(M)$>$1.25 $M_{\odot}$. The lighter the observed star, the greater the mass difference, for a given value of $f$.

\begin{figure}
    \centering
	\includegraphics[width=0.46\textwidth]{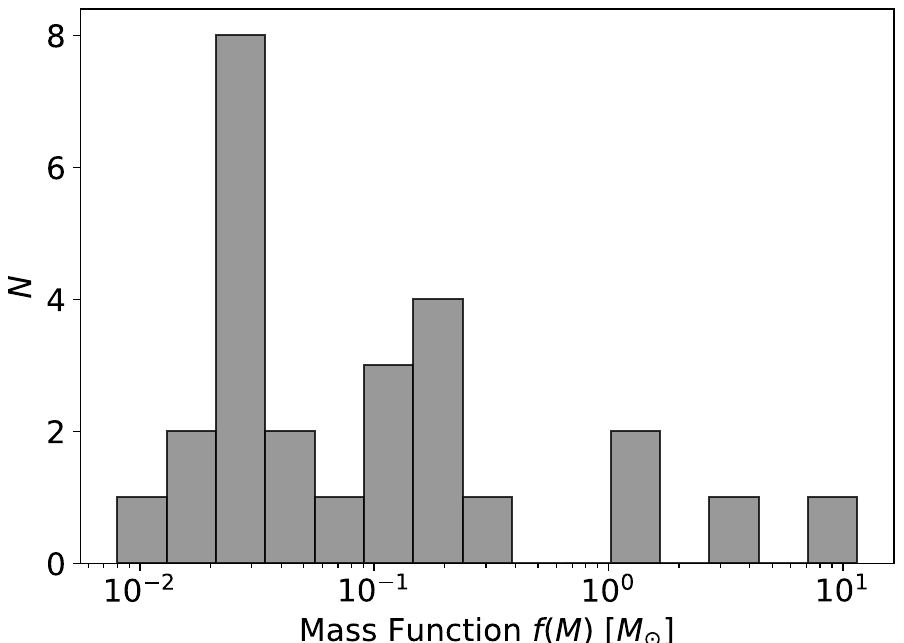}
    \caption{Mass function f(M) distribution of the 27 well constrained binaries in NGC 1850. Most of the binaries are characterized by f(M)$\ll 1\,M_\odot$ but a tail at values above 1.25$M_{\odot}$ is observed.}
    \label{fig:MF}
\end{figure}

\begin{figure*}
    \centering
	\includegraphics[width=0.75\textwidth]{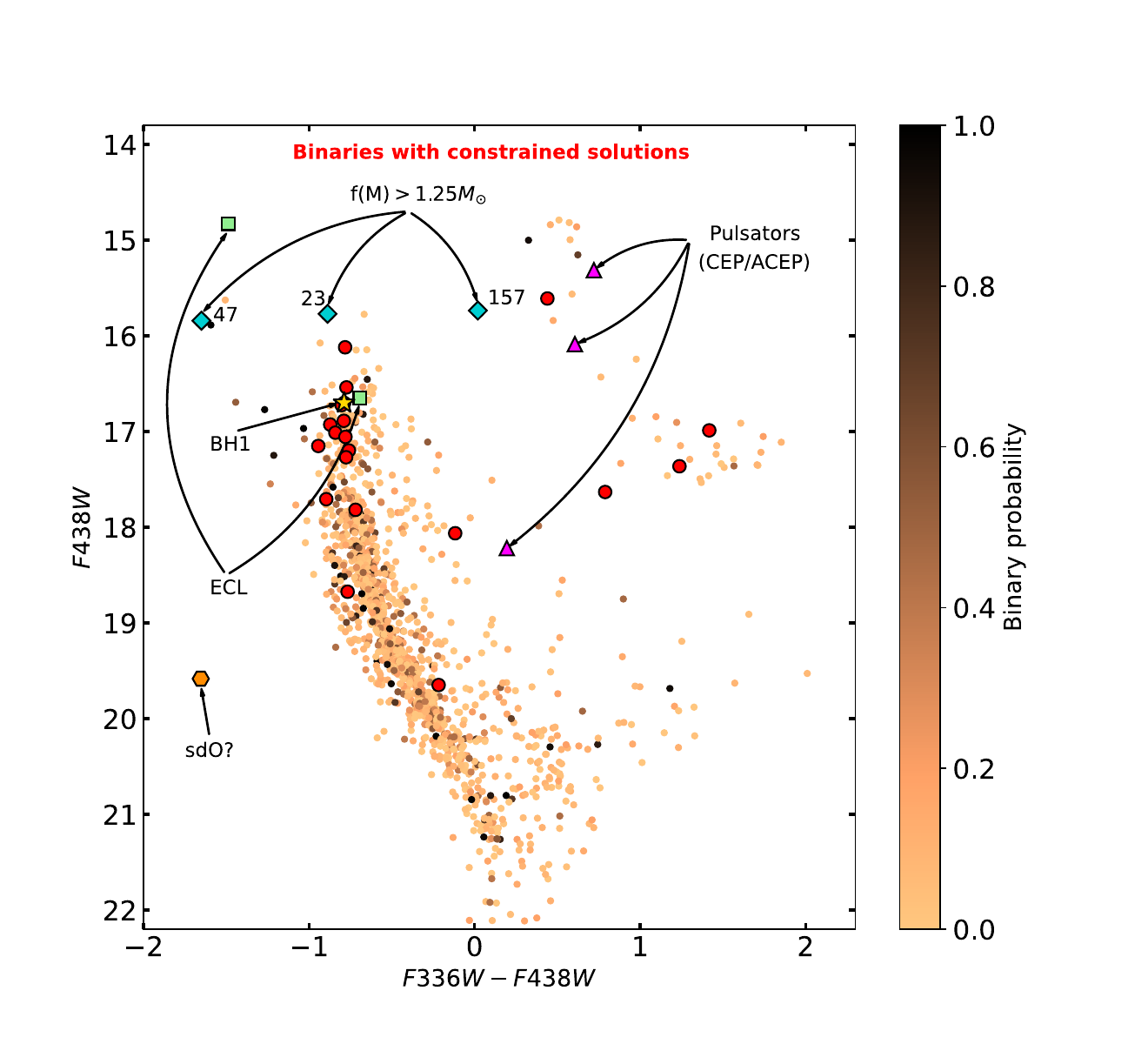}
    \caption{(F438W, F336W-F438W) CMD of NGC 1850 as in Figure \ref{fig:cmd}, where all binary systems with constrained solutions from \textsc{The Joker} (i.e. orbital properties, P, K, e etc.) are highlighted as large red dots. Different types of stars/binaries are also highlighted with different symbols and colours, as labeled in the figure (e.g. Eclipsing binaries (ECL) and Pulsators from the OGLE-IV survey, systems with massive companions (f(M)$>$1.25 $M_{\odot}$) and a possible subdwarf O-type star). Systems with massive companions are also labeled individually, as they will be discussed in detail later in the text.}
    \label{fig:cmd_constrained}
\end{figure*}

Figure \ref{fig:MF} shows the histogram of the distribution of the binary mass function for our sample of 27 constrained binaries. As can be seen, most of the stars exhibit a low binary mass function (f(M)$\ll$1 $M_{\odot}$). This is consistent with what we expect from the visual inspection and analysis of their spectra. In fact, they all appear to be SB1 binaries, i.e. binaries in which the flux contribution of the observed star is significantly larger than that of the companion, which is not detected. In most cases, this is due to the fact that the invisible source is so faint (therefore of low mass) that its contribution is negligible. Looking at the distribution, however, we note that there are four systems which have a mass function exceeding 1.25 $M_{\odot}$, with f(M) values between 1.25 $M_{\odot}$ and $\sim$10 $M_{\odot}$. With significantly massive companions, these systems are the same ones we identified in Figure \ref{fig:distributionK} and are very interesting systems as they are the most likely candidates to host a compact dark object, such as a neutron star (NS) or a BH. We sound a note of caution here that there may be several reasons why the contribution of a second bright star might not be detected in stellar spectra. For example, if it rotates so fast that all the spectral lines become very broad (e.g. LB-1, \citealt{Shenar2020LB1} and HR 6819, \citealt{Bodensteiner2020HR}) or if it has an accretion or decretion disk that partially blocks its light. Further observations are needed to characterize the nature of the unseen sources in the binaries with f(M)$>$1.25 $M_{\odot}$, however a compilation of all the information we have is presented later in the text.
The main results of this work can be well summarized by modifying the (F336W-F438W, F438W) CMD of NGC 1850 in Figure \ref{fig:cmd} to include all the information we obtained from the analysis performed with \textsc{The Joker}, as well as the comparison with other photometric catalogs in the literature. Figure \ref{fig:cmd_constrained} shows all stars observed in the MUSE FOV for multiple epochs and for which we have estimated a probability to be in binary systems. Dark colors indicate a high probability that they are in binaries, while light orange colors suggest that they are single stars. The 27 binaries with constrained orbital properties are shown as large red dots and are found mostly in the upper part of the main-sequence, being bright. The binary probability is larger than 90\% ($p>0.9$) for all constrained binaries, except for the one with bimodal solution. Figure \ref{fig:cmd_constrained} also highlights different types of stars/binaries to emphasize some classes of objects that we have identified in our sample. Eclipsing binaries are shown as large green squares (including star \symbol{35}51, that we have classified as a member of NGC 1850B) while binaries with massive companions (f(M)$>$1.25 $M_{\odot}$) are indicated by large cyan diamonds. The large yellow star denoting the BH candidate BH1, also belongs to the latter category (its properties are discussed in \citealt{Saracino2022,saracino2023}). The orange hexagon identifies a binary system that could harbor a stripped subdwarf O (sdO) star candidate. Finally, large pink triangles highlight the variable stars (mainly Cepheids) found in NGC 1850, all of which are located in the evolved portion of the CMD. Each of these object classes will be discussed in the next section, where more details will be provided.
\vspace{-0.5cm}
\section{Interesting objects in NGC 1850}
\label{sec:individual_systems}
The detailed analysis of the binary content of NGC 1850 has shown the presence of a few very interesting binary systems, of very different natures. Here we discuss those cases separately.
\vspace{-0.5cm}
\subsection{Binaries with massive companions}
The orbital properties derived by \textsc{The Joker} suggest that a sample of 4 out of 27 binaries have a binary mass function f(M)$>$1.25 $M_{\odot}$. They are the binaries with HST ID \symbol{35}23, \symbol{35}47, \symbol{35}157 and \symbol{35}224. The latter has been already extensively discussed in \citet{Saracino2022,saracino2023} and will not be mentioned here. Instead we focus on the others later in the text.

The first interesting property we note is when a system has a value of f(M)$>$1.25 $M_{\odot}$. In the context of NGC 1850, this tells us that the component of the binary we do not see is more massive than the star we observe. In the first place, this finding only means that the mass ratio q of the binary is above 1, if it is defined as q = $M_{\rm unseen}$/$M_{\rm visible}$, but it does not provide any information about the nature of the object we do not observe. 

Indeed, a source could be massive but not observed because: 1) it is rotating so fast that the blurring of the absorption lines hamper the lines from the two components to be observed; 2) there is an accretion disk which blocks a significant fraction of its light, making it difficult to detect; 3) it is a dark object, such as a NS or a BH, so it emits no observable light.

We have looked more into the binary system with the highest binary mass function, f(M)=10.53 $M_{\odot}$ or 14.46 $M_{\odot}$ and we report the details below.
\vspace{-0.5cm}
\subsubsection{Star \symbol{35}47}
For star \symbol{35}47 we have a total of 25 radial velocity measurements, spanning a range of more than three years. In particular, 9 observations come from the WiFeS instrument (WAGGS survey), while the remaining 16 epochs come from MUSE observations. This star has been observed only in one of the two MUSE pointings of NGC~1850, i.e. the central pointing, meaning that we were able to extract an high S/N spectrum for every image secured. 

\begin{figure*}
    \centering
    \includegraphics[width=0.95\textwidth]{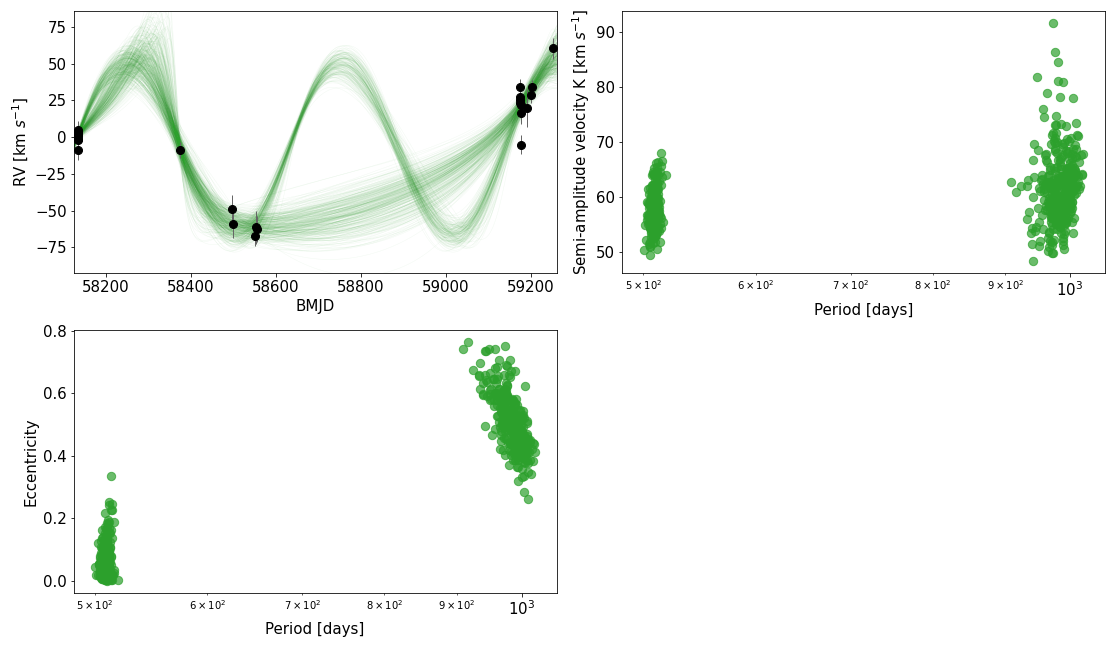}
    \caption{{\it Left:} Radial velocity curve (black data points) of star \symbol{35}47 in NGC 1850. The green curves are the possible orbital solutions determined by \textsc{The Joker} for this star. {\it Right:} Period [P] - Semi-amplitude velocity [K] plot of these samples. {\it Bottom:} Period [P] - Eccentricity [e] plot of these samples. The orbital properties of this system are well constrained, i.e. it is a unimodal solution but given the high measured binary mass function, it is worth analyzing both clusters of solutions separately.}
    \label{fig:star47}
\end{figure*}

\begin{figure}
    \centering
    \includegraphics[width=0.48\textwidth]{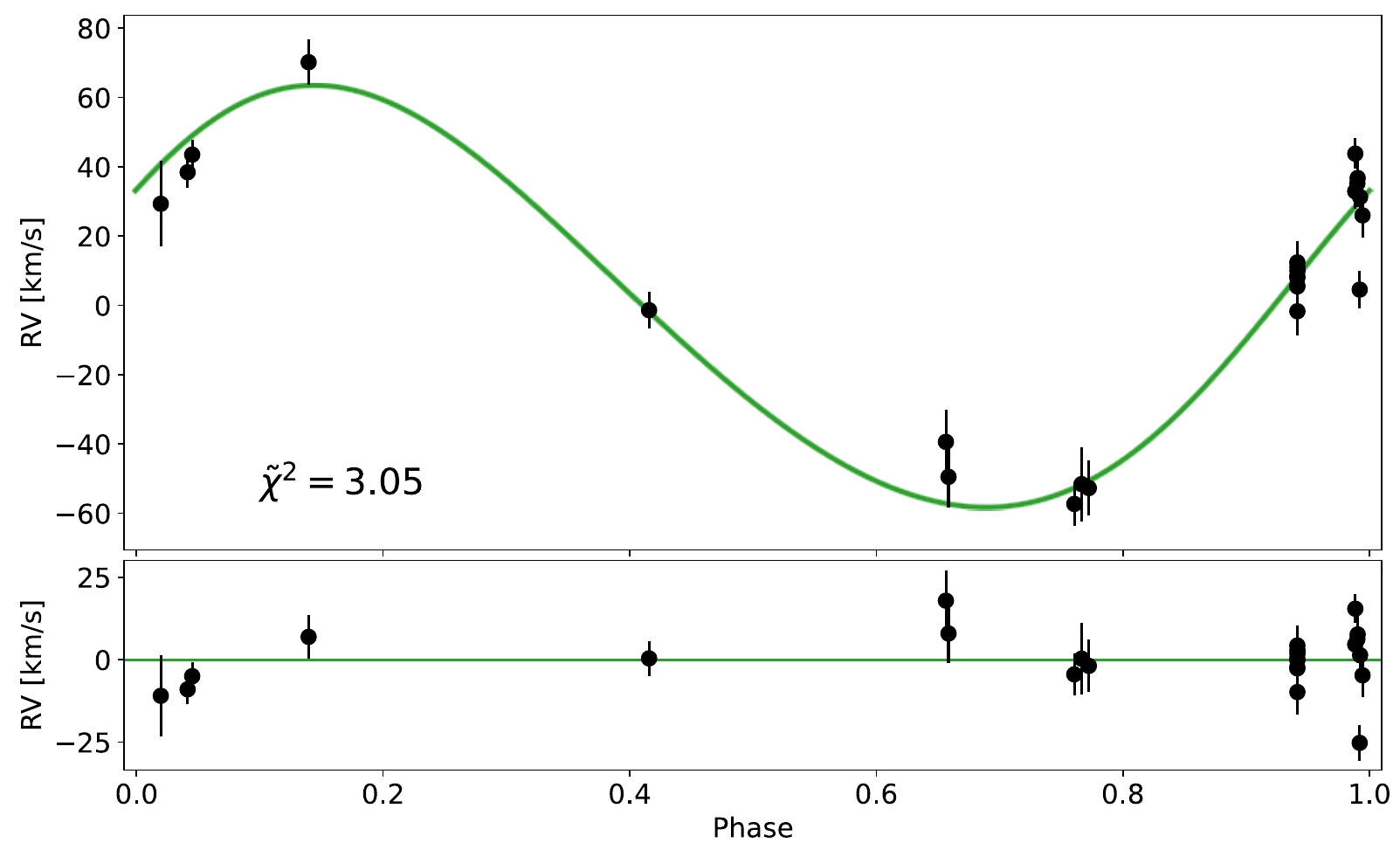}
    \caption{{\it Left:} Phase-folded radial velocity curve (black data points) of star \symbol{35}47 in the NGC 1850 sample, by assuming an orbital period P=508.4 d. The green curve is the median orbital solution determined by \textsc{The Joker} for this star when considering only samples with period P$<$600 d. {\it Bottom:} Residuals of the comparison between the data and the best-fit orbital solution, where the reduced $\chi^2$ of the fit is also shown. The combination of orbital properties (P, e, K) suggests a mass function for \symbol{35}47 and for this cluster of solutions of f(M)=10.53 $M_{\odot}$.}
    \label{fig:star47_phase1}
\end{figure}

\begin{figure}
    \centering
    \includegraphics[width=0.48\textwidth]{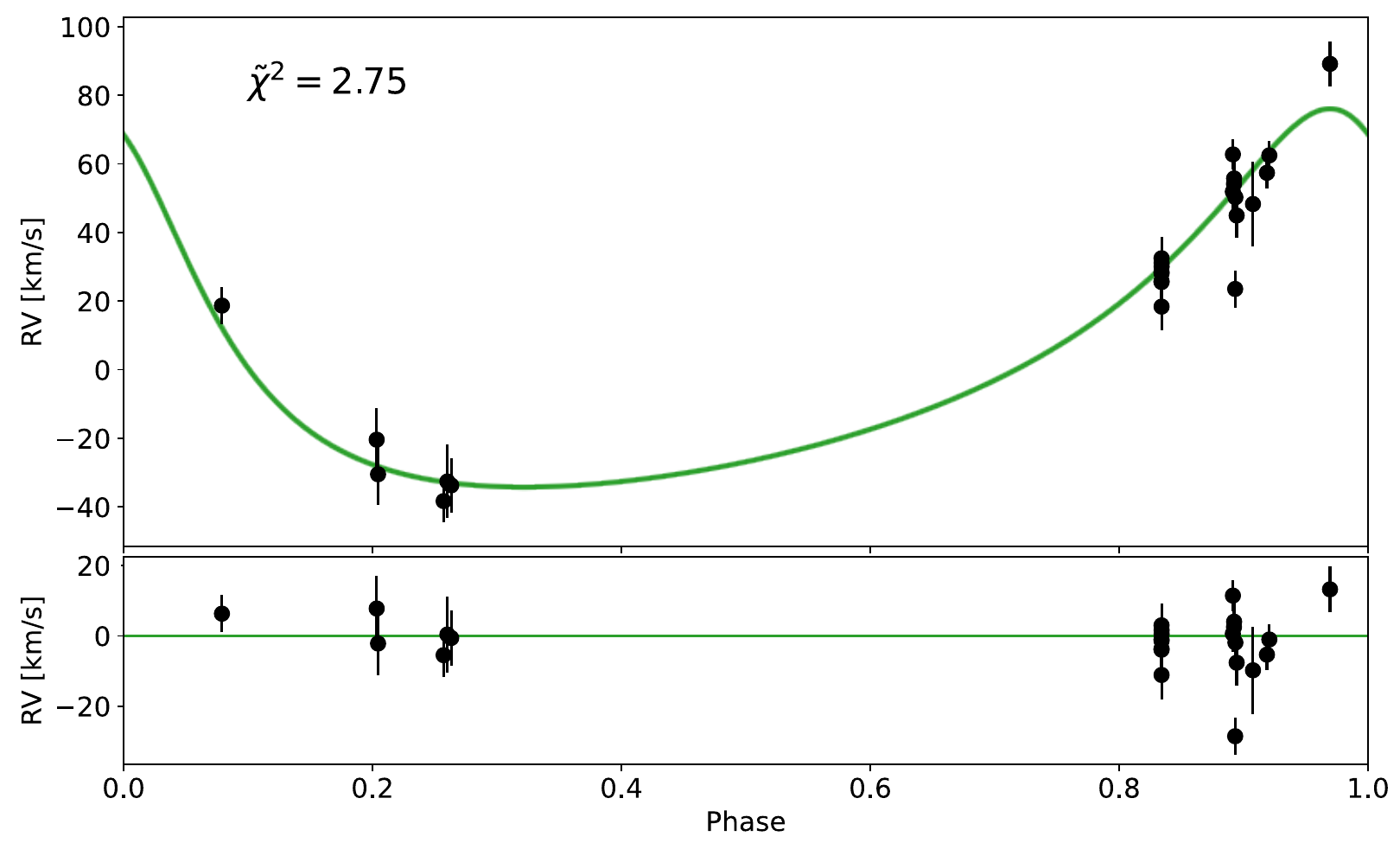}
    \caption{{\it Left:} Same as for Figure \ref{fig:star47_phase1}, but assuming an orbital period of P=984.3 d, obtained as the median period when only samples with P$>$600 d (second cluster of solutions) are considered. The combination of orbital properties (P, e, K) suggests a mass function of f(M)=14.46 $M_{\odot}$.}
    \label{fig:star47_phase2}
\end{figure}

According to the criterion we have adopted so far in the paper, star \symbol{35}47 has a unimodal solution, however, when looking closer at the posterior samples, \textsc{The Joker} finds for the system two possible orbital solutions, with periods P=508.4 d and P=984.3 d. These solutions, shown in Figure \ref{fig:star47}, are very well defined when WiFeS and MUSE observations are analyzed together, however they are already well recognizable when only the MUSE data are used. The phase-folded radial velocity curve of star \symbol{35}47 using the first and the second best-fit solution are presented in Figures \ref{fig:star47_phase1} and \ref{fig:star47_phase2}, respectively, where also residuals are shown. The two solutions are equally good, as demonstrated by the residuals as well as the reduced $\chi^2$ value reported. 
It is clear that star \symbol{35}47 belongs to the class of long-period binaries. There is no evidence to suggest that any short-period solutions can be missed. Furthermore, MUSE observations were designed to be extremely sensitive to binaries with orbital periods of the order of days, so given the available measurements, if the system was a short period binary, we would definitely have detected it. 

\begin{figure}
    \centering
    \includegraphics[width=0.43\textwidth]{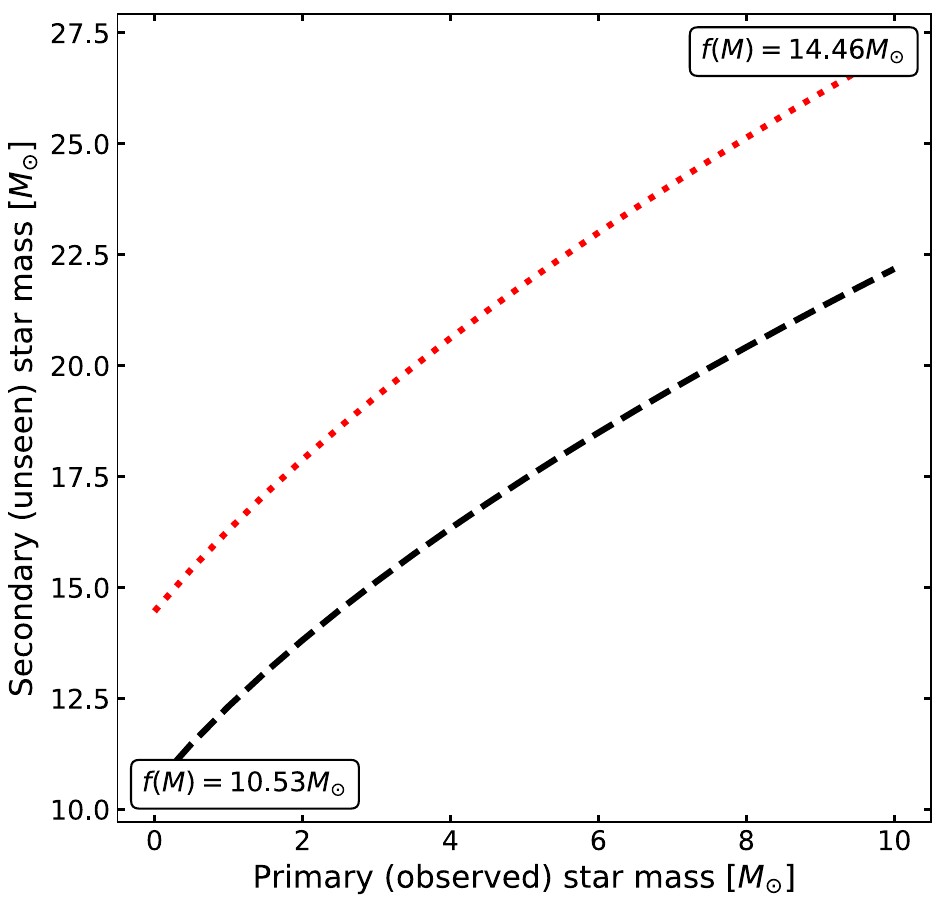}
    \caption{Primary (observed) mass versus secondary (unseen) mass for star \symbol{35}47, according to the two best-fit orbital properties presented in Figures \ref{fig:star47_phase1} and \ref{fig:star47_phase2}. In particular, the black dashed line represents the mass function f(M) for a period P=508.4 d, while the red dotted line identifies the mass function f(M) for a period P=984.3 d. In both cases the inclination $i$ of the binary is assumed to be of 90$^\circ$. The true orbital properties of star \symbol{35}47 are not yet fully constrained but in either case, the mass function is so high that for a reasonable mass of the visible star (up to 5 $M_{\odot}$ if it belongs to NGC 1850 and possibly up to 10 $M_{\odot}$ if it belongs to the very young cluster NGC  1850B), the companion, primary star has to have a mass of at least 10 $M_{\odot}$, and it cannot be anything other than a dark object, such as a BH.}
    \label{fig:star47_MF}
\end{figure}

\begin{figure*}
    \centering
    \includegraphics[width=0.7\textwidth]{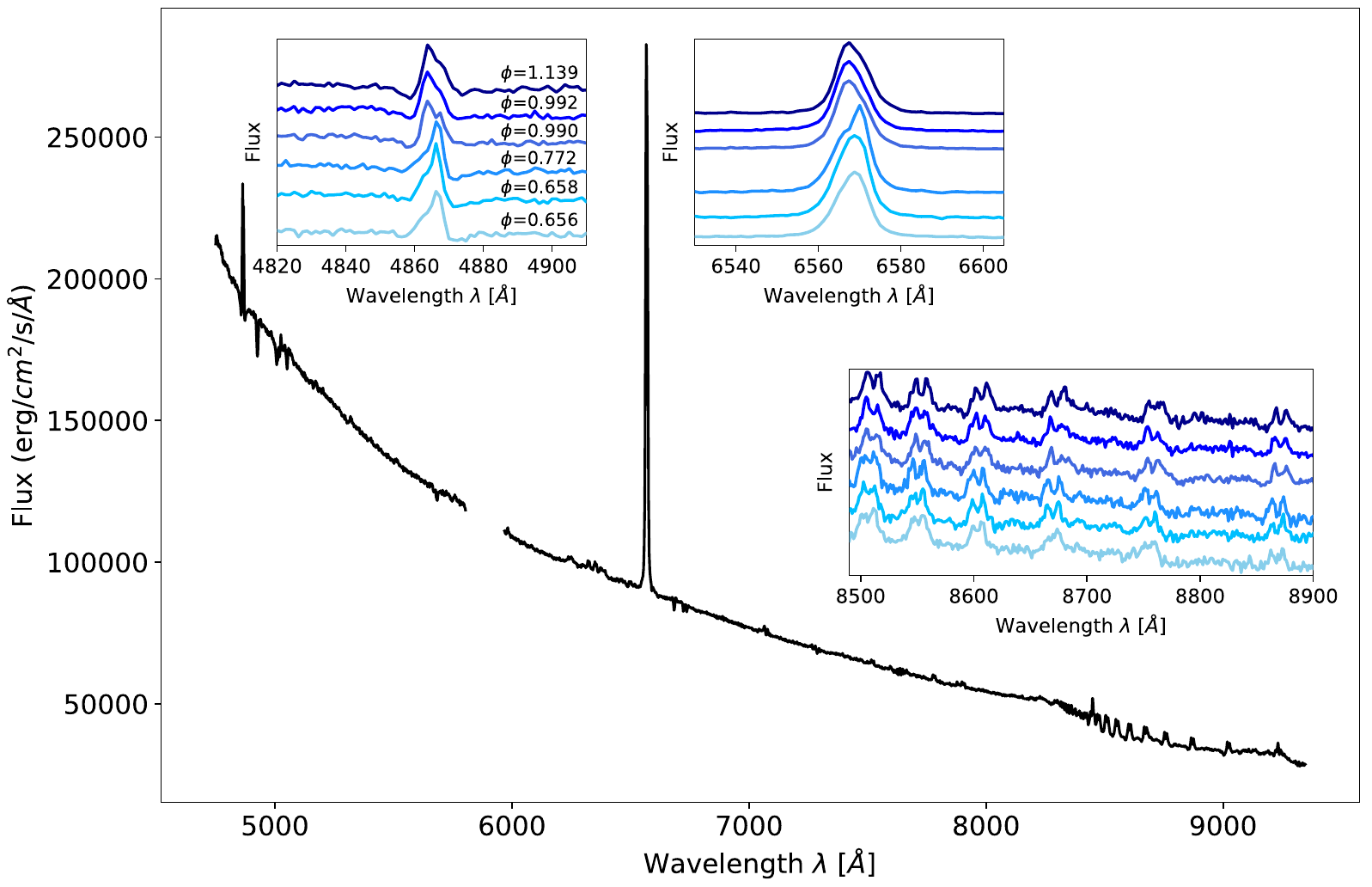}
    \caption{The combined, rest-framed, spectrum of star \symbol{35}47, over the full MUSE wavelength range (black line), where strong H$\beta$ and H$\alpha$ emission are observed. Three insets are shown in the same plot, with a zoom-in in three different portions of the spectrum, H$\beta$, H$\alpha$ and the Paschen series, respectively. The insets also contain a sample of spectra of star \symbol{35}47, taken at different times, and distributed in colours from light to dark for increasing phases $\phi$, as also labelled in the figure. The $\phi$ values were determined by using a period P=508.4 d, but we have verified that the trend still holds for the second period P=984.3 d. As can be observed for the H$\beta$ and H$\alpha$ lines, the shape of the emission line clearly changes with phase. The Paschen series instead present a set of double-peaked emission lines which move coherently as a function of time (phase). This is a clear evidence for the presence of an accretion/decretion disk in the system, orbiting around the visible star.}
    \label{fig:spectra47}
\end{figure*}

As presented in Table \ref{tab:properties}, both orbital solutions derived by \textsc{The Joker} imply a binary mass function f(M) well above 1.25 $M_{\odot}$. In particular, f(M)=10.53 $M_{\odot}$ for the P = 508.4 d solution and f(M)=14.46 $M_{\odot}$ for the P = 984.3 d solution. This means that the unseen companion in this system is considerably more massive than the star we observe and this is reported in Figure \ref{fig:star47_MF}, where the two curves (black-dashed and red-dotted), computed based on the measured mass functions, show how massive the unseen component of the system would be as a function of the assumed mass of the visible component. On the x-axis, a range 0-10 $M_{\odot}$ is presented, indicating all the possible masses the observed star can assume (0 $M_{\odot}$ on one extreme, 10 $M_{\odot}$ on the other, if the star belongs to the 5 Myr-old cluster NGC 1850B). This plot seems to suggest that the system containing star \symbol{35}47 needs to host a massive BH candidate, in order to explain such a trend.
While the available data are not sufficient to fully understand the nature of the system, we have analyzed the spectra of this peculiar system, also applying the disentangling technique (see \citealt{saracino2023} and references therein for details on the methodology) and we summarize below all the information acquired:
\begin{itemize}
\item The spectra of star \symbol{35}47 show prominent emission in H$\alpha$ and H$\beta$, as well as a double-peaked emission in the Paschen series. The shape of the Balmer lines varies as a function of the orbital phase, while the shape of the Paschen series do not (see the insets in Figure \ref{fig:spectra47}).
\item The HeI line at $\sim$4920 \AA, and the Paschen lines are moving in phase, while the Balmer lines (H$\alpha$ and H$\beta$) result in close-to-constant radial velocities.
\item The disentangling of the He I line and the Paschen lines yields no notable features for the presence of a secondary in the system, showing a close-to-flat spectrum. On the other hand, the disentangling of the Balmer lines identifies two components: a clear single-peak emission component, superimposed on a double-peaked emission component. The contribution in flux of the single-peak emission component is not negligible and we estimated it to contribute $\approx$30-40\% on the entire flux in H$\alpha$ and H$\beta$.
\item We measured the rotational velocity of the observed star from the He I line at $\sim$4920 \AA\, in the MUSE spectrum and we found a value close to 300 $\,{\rm km\,s^{-1}}$. The observed star is a rapid rotator, and its rotational velocity is close to the break-up velocity observed for other stars in NGC 1850 (the detailed analysis is presented in \citet{Kamann2023} and their Figure 11).
\end{itemize}
By combining all the information, it seems clear that the spectrum of the visible star is rather similar to that of a Be-type star (also corroborated by the position of star \symbol{35}47 in the region of the CMD where blue stragglers are located, see Figure \ref{fig:cmd_constrained}), surrounded by a decretion disk.
Another disk also appears to be present in the system, around the massive and unseen object. It is responsible for the single-peaked emission in the Balmer lines. The fact that this emission is observed at rest is in line with the expectations, because the center of mass of a system made up of two components in which one is significantly more massive than the other coincides almost perfectly with the position of the massive source itself.
Much information has been recovered from the MUSE spectra, but the nature of the unseen component is still unclear. One of the most plausible interpretations, however, could be that star \symbol{35}47 is part of a binary or a tertiary system. If it is a binary, a Be-type star with a decretion disk is orbiting with a long period around a massive object (e.g. a BH candidate, given its mass), which has its own accretion disk. In a tertiary configuration instead, star \symbol{35}47 would orbit, with a short period, around a donor (low-mass and faint) star from which it has gained most of the mass, and the whole binary orbits around a massive object, such as a BH candidate, on a long orbit. The accretion disk may have been generated by accretion of material onto the BH e.g. by strong winds coming from star \symbol{35}47.
\vspace{-0.5cm}
\subsubsection{Star \symbol{35}23 and star \symbol{35}157}
Star \symbol{35}47 is not the only peculiar source observed in the inner parts of NGC 1850. In fact, there are two other sources, star \symbol{35}23 and star \symbol{35}157, which have similar properties to star \symbol{35}47. For example they have comparable luminosity in F438W (see Figure \ref{fig:cmd_constrained}), belong to the class of long-period binaries and are characterized by a binary mass function f(M)$>$1.25 $M_{\odot}$. Unlike star \symbol{35}47, however, the values we are dealing with here are much less extreme and that is why we treat them together.

\begin{figure*}
    \centering
    \includegraphics[width=0.95\textwidth]{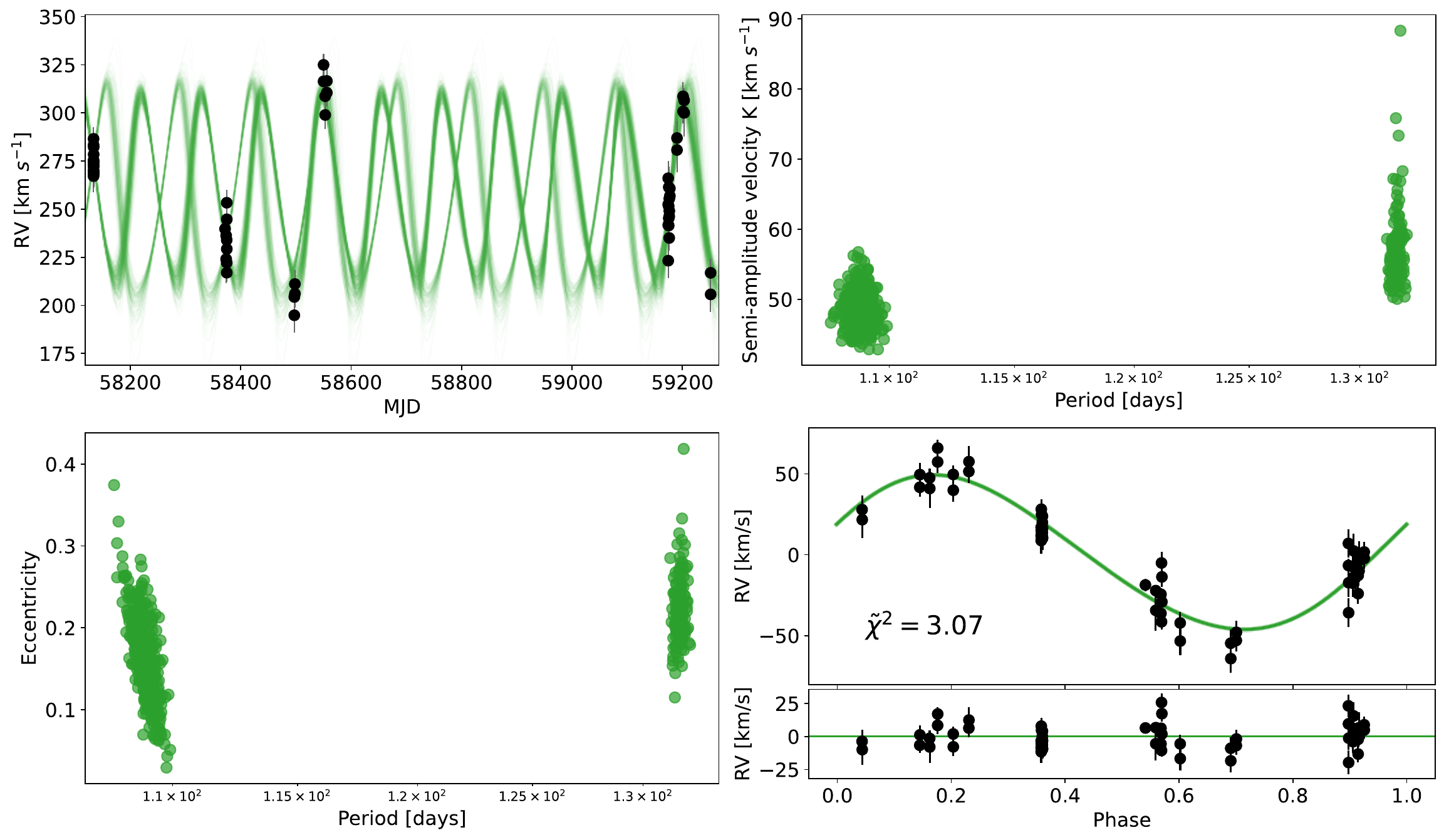}
    \caption{{\it Top-left:} Radial velocity curve (black data points) of star \symbol{35}23 in the NGC 1850 sample. The green curves are the possible orbital solutions determined by \textsc{The Joker} for this star. {\it Top-right:} Period [P] - Semi-amplitude velocity [K] plot of these samples. {\it Bottom-left:} Period [P] - Eccentricity [e] plot of these samples. The orbital properties of this binary are well constrained according to the methodology proposed by \citet{Giesers2019}, i.e. unimodal solution. However, two main clusters appear, with the most populated one located around period P=109.1 d. {\it Bottom-right:} Phase-folded radial velocity curve (black data points) of the binary, by assuming the best fit orbital period P = 109.1 d. The reduced chi-square of the fit is also reported.}
    \label{fig:star23}
\end{figure*}

Star \symbol{35}23 is located in the brightest part of the main-sequence of NGC 1850, and has 58 radial velocity measurements in total, the first 26 from the WAGGS spectra, while the remaining 32 from MUSE. This star falls in the overlapping region between the two MUSE pointings, so it has as many epochs as possible (see Figure \ref{fig:epochs}). Considered well constrained (i.e. it has a unimodal solution) based on the criterion in \citet{Giesers2019}, this star has an orbital period P=109.1 d according to \textsc{The Joker}. Figure \ref{fig:star23} shows two main clusters of solutions for the star, but the one around P$\sim$130 d is significantly less populated than the other. That is the reason why we do not treat the two clusters separately, but take them both into account when calculating the median and the uncertainty associated with its orbital parameters (see Table \ref{tab:properties}). In the bottom-left panel of the figure, we also show the phase-folded radial velocity curve of the binary, which fits well the behavior of the observed velocities. Star \symbol{35}23 has an estimated f(M)=1.26 $M_{\odot}$.

\begin{figure*}
    \centering
    \includegraphics[width=0.95\textwidth]{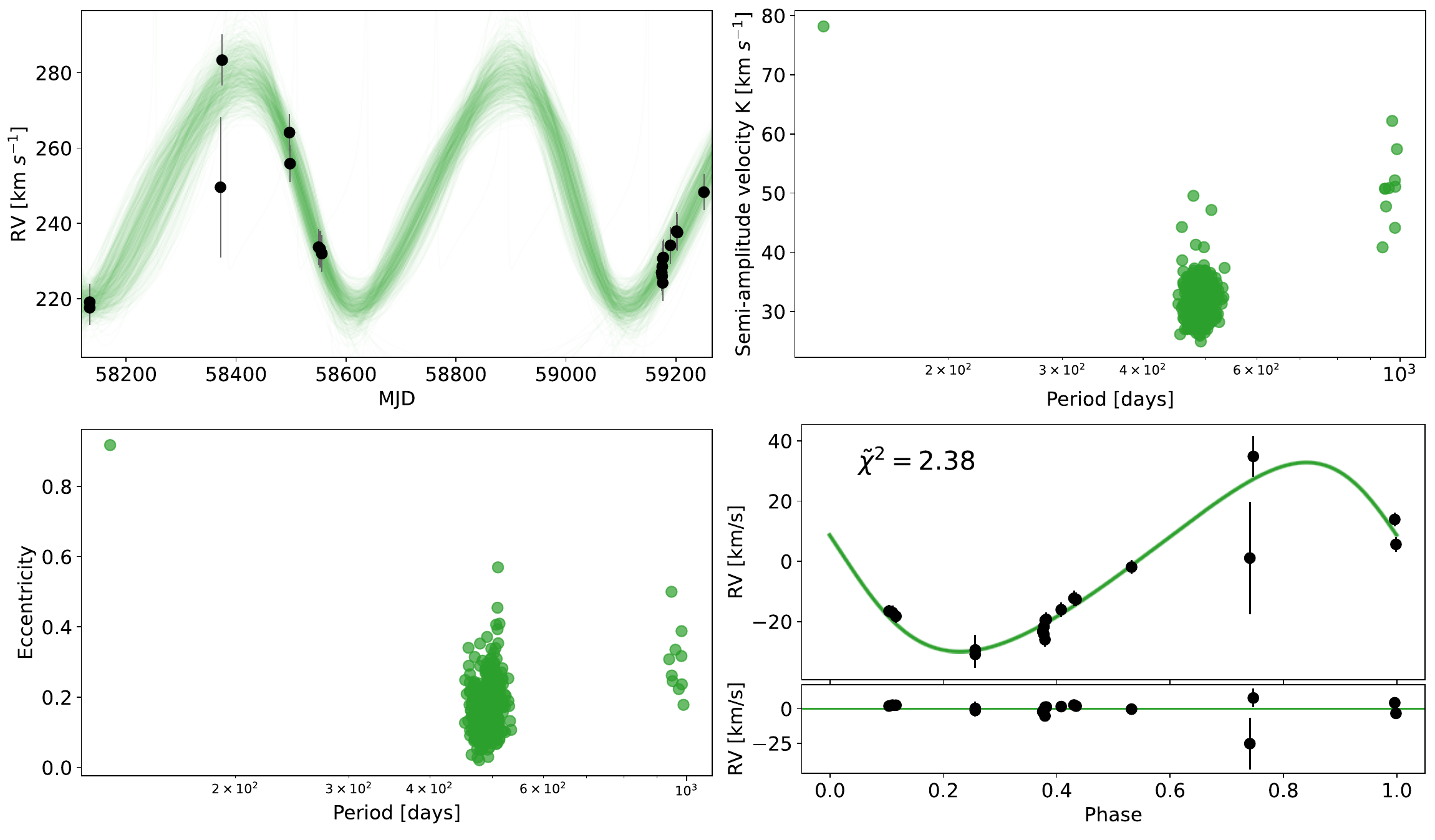}
    \caption{As in Figure \ref{fig:star23} but for star \symbol{35}157. The best fit orbital period in this case is P = 491.0 d.}
    \label{fig:star157}
\end{figure*}

On the other hand, star \symbol{35}157 is rather off the main sequence, located at intermediate colors between the unevolved and the evolved sequences. We constructed the radial velocity curve using 4 observations from WAGGS and 16 observations from MUSE. The best-fit orbital solution derived by \textsc{The Joker} for this system has a period of 491.0 d, as can be seen in Figure \ref{fig:star157}, and its phase-folded radial velocity curve also well reproduces the observations (see the bottom-right panel of the same figure). For star \symbol{35}157 we obtained a binary mass function of 1.45 $M_{\odot}$.
Their spectra do not show striking features like that of star \symbol{35}47 (with the exception of the non-standard Paschen lines in the spectra of star \symbol{35}23), so we have not performed further investigations. The only information we have so far is that their invisible companions must be quite massive. However, the latter could be either dark remnants such as NSs and BHs, or stellar components which due to some properties of the system are not easy to detect.
\vspace{-0.5cm}
\subsection{Star \symbol{35}2413 - O-type subdwarf candidate}
Star \symbol{35}2413 is located in a particular region of the CMD in Figure \ref{fig:cmd_constrained}. It has a magnitude of F438W=19.58 but it is significantly bluer than the main sequence of NGC 1850 (by $\sim$1 dex in F336W-F438W). Furthermore, it has more than 95\% probability of being a binary system. The star is relatively faint, so only 5 radial velocity measurements from MUSE survived the quality cuts we imposed to obtain the final catalog of radial velocities in the cluster. For the same reason, no WAGGS spectra are available. The number of radial velocity measurements therefore is not sufficient for \textsc{The Joker} to constrain the orbital properties of the system. 

\begin{figure*}
    \centering
	\includegraphics[width=\textwidth]{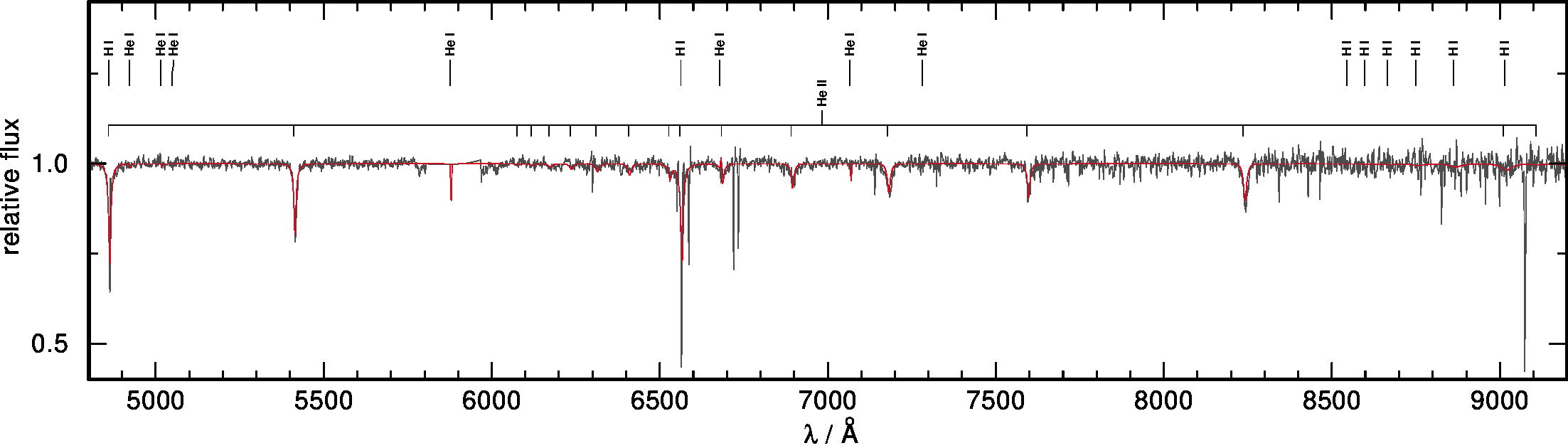}
    \caption{Spectrum of the sdO star candidate in NGC 1850 (in black) with the best-fit template overplotted in red. The resulting stellar parameters obtained from the comparison with the best-fitting stellar template are T\textsubscript{eff} = 50,653 $\pm$ 791 K, log(g) = 5.20 $\pm$ 0.04, and log(He/H) = -0.58 $\pm$ 0.04 (number fraction).}
    \label{fig:star2413}
\end{figure*}

The information we have for star \symbol{35}2413 are limited, but its spectra are very interesting. No metal lines can be detected, but the He II 4859\,\AA / H$\beta$ blend, the He II 6560\,\AA / H$\alpha$ blend, several other pure He II lines and a weak but clearly detectable line of He I at 7065\,\AA. An example is shown in black in Figure \ref{fig:star2413}. 

\begin{figure}
    \centering
	\includegraphics[width=0.46\textwidth]{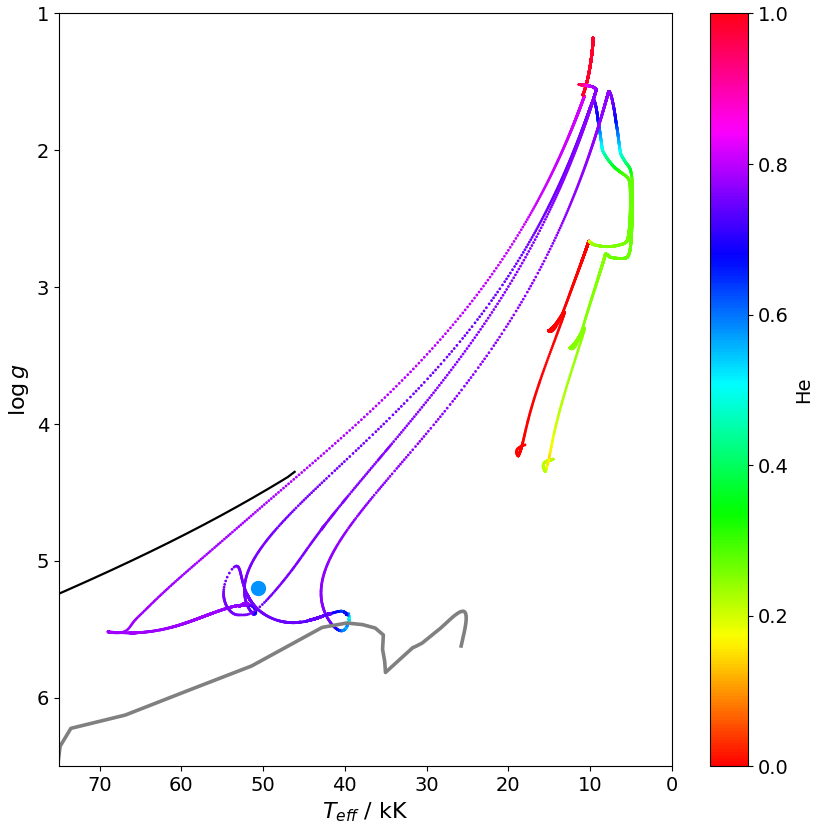}
    \caption{Theoretical T\textsubscript{eff} vs log(g) tracks of stripped hot subdwarf stars from \citet{gotberg2018} corresponding to initial masses of 6.3 $M_{\odot}$ and 4.04 $M_{\odot}$ and final masses of 1.01 $M_{\odot}$ and 0.82 $M_{\odot}$, respectively), colour-coded for different He abundances. The blue point indicates the location of our source (star \symbol{35}2413) in this parameter space, with the color linked to its He abundance. Its position appears to be consistent with it being a stripped star. The black line represents a 0.525 $M_{\odot}$ post-AGB track from \citet{2010ApJ...717..183R} while the grey line shows a 0.471 $M_{\odot}$ post-extreme horizontal branch track from \citet{1993ApJ...419..596D}.}
    \label{fig:HRD}
\end{figure}

Interestingly, this spectrum appears strikingly similar to those presented in \citet{gotberg2018}, which are the typical spectra of B- and O- type (sdB-O) subdwarf stars (i.e. very hot stars that have been stripped of most of their material due to mass transfer and are now substantially left with a He core). These stars go through a short-lived phase \citep{irrgang2020}, in which they are quite bright, and end up in this magnitude range once they are relaxed and have reached the stripped phase. 
To investigate whether this might be a stripped sdO-type star, we adopted very hot stellar templates \citep{2016A&A...587A.101R} from which we estimated effective temperature T\textsubscript{eff}, surface gravity log(g) and He abundance log(He/H) of star \symbol{35}2413 via full-spectrum fitting. The stellar parameters obtained from the comparison with the best-fitting stellar template (shown in red in Figure \ref{fig:star2413}) are T\textsubscript{eff} = 50,653 $\pm$ 791 K, log(g) = 5.20 $\pm$ 0.04, and log(He/H) = -0.58 $\pm$ 0.04 (number fraction). When comparing these numbers with theoretical tracks of stripped hot subdwarf stars from \citet{gotberg2018}, the location of our source in NGC 1850 matches the predictions quite well (shown in Figure \ref{fig:HRD}). This is a strong indication that star \symbol{35}2413 is indeed a stripped sdO star candidate orbiting a companion. We note that most of the known sdO stars are located at higher T\textsubscript{eff} and surface gravity \citep{2022A&A...662A..40C}. The T\textsubscript{eff} and log(g) of star \symbol{35}2413 seem a bit too low for a post-extreme horizontal branch star \citep{1993ApJ...419..596D} and also a post-asymptotic giant branch nature seems unlikely since at T\textsubscript{eff}= 50,653 K, the log(g) would be required to be lower \citep{2010ApJ...717..183R}. Interestingly, the \citet{gotberg2018} tracks actually predict that - for this mass - the companion should dominate the optical spectrum. In conclusion, more data is needed to characterize the evolutionary history of this star.
\vspace{-0.5cm}
\subsection{Pulsating stars}
\label{pulsators}
As mentioned above, the aim of this work is to detect radial velocity variations due to the motion of a star around a companion, rather than an intrinsic variability physically linked to the star itself (i.e. pulsations). These two classes of objects are in principle indistinguishable, as they manifest themselves in a similar manner, however there are ways to discriminate between them. 

There are compilations of variable stars (e.g. RR Lyrae or Cepheids) detected via photometric monitoring with dedicated instruments for the Milky Way, as well as for the LMC and SMC. The Optical Gravitational Lensing Experiment (OGLE, \citealt{OGLE1992}), for example, is one of the surveys that has identified and characterized the largest number of variable systems in the last decades, even in star clusters. 

To clean up our sample of these objects, we compared our final catalog of binaries with the OGLE-IV release. We found a list of four sources in common between the two catalogs, including three Cepheids (star \symbol{35}96 or CEP-1082; star \symbol{35}306 or CEP-1078 and star \symbol{35}506 or CEP-1080) and one Anomalous Cepheid (star \symbol{35}1916 or ACEP-024). Of those, 3 out of 4 have orbital properties constrained by \textsc{The Joker}. They are not included in the 27 systems with constrained orbits and are not in Table \ref{tab:properties} as they are not binary systems. 

The period derived from \textsc{The Joker} for these stars is in excellent agreement with the pulsational period derived from the OGLE-IV light curves, confirming the accuracy and reliability of our radial velocity measurements and results. The fourth pulsator in the sample (star \symbol{35}96) has a constrained solution according to \textsc{The Joker} (P$\approx$ 5.5 d), but the orbital period does not match with the one provided by OGLE-IV (P=7.86 d). We have used both the \textsc{Spexxy} and CC radial velocities for this star and the solutions we got were always the same, hence we are not able to explain where the discrepancy comes from.
\vspace{-0.5cm}
\subsection{Eclipsing binaries}
By comparing OGLE-IV with our catalog of binaries in NGC 1850, we identified five other sources, this time true radial velocity variables, which are classified as eclipsing binaries (ECL), i.e. binaries with an inclination such that as one star passes in front of the other, it creates a dip in the light curve of the system. ECL light curves are all very different from each other and their shape really depends on the configuration of the system, as well as the mass ratio of the components involved.  
Only 3 out of 5 have unimodal solutions from \textsc{The Joker} (star \symbol{35}51 or ECL-08397; star \symbol{35}191 or ECL-08469 and star \symbol{35}224 or ECL-29851). The latter, also called NGC1850 BH1, was the focus of two dedicated works \citep{Saracino2022,2022MNRAS.511L..24E} which demonstrated that the source is not an eclipsing but an ellipsoidal binary. In all three cases, the orbital period measured by \textsc{The Joker} almost exactly matches that listed by OGLE-IV, providing independent confirmation that they are indeed binaries. Given its spatial position in the MUSE FOV, there is a high probability that star \symbol{35}51 is not member of the main cluster NGC 1850, but of the much younger cluster NGC 1850B. An example of an eclipsing binary is shown in Figure \ref{fig:ecl} of Appendix \ref{app:extra}. The phase-folded radial velocity curve of \symbol{35}191 as derived from MUSE (using P=6.46 d) is presented in the left panel, while the phase-folded light curve by adopting the same orbital period is in the right one. 

Eclipsing binaries have huge potential. Indeed, when both radial velocity curves and light curves are available for the same binary system, the degeneracy between the inclination of the system and the mass ratio of the two components can be broken and these properties can be conveniently inferred. By adopting $i$ = 90$^\circ$ (given the observed eclipses), and visible star masses inferred from the comparison to best-fit isochrones ($M_{1}$ $\sim$ 14.79 $M_{\odot}$ for star \symbol{35}51, since it belongs to NGC 1850B and $M_{1}$ $\sim$ 4.98 $M_{\odot}$ for star \symbol{35}191, member of NGC 1850), we used Equation \ref{eq:fm2} to estimate a value of the mass ratio q for the two eclipsing binaries in our sample. In particular we derived q = 0.14 for star \symbol{35}51 and q = 0.45 for star \symbol{35}191. These values are an approximation, since they are based on assumptions that may not be entirely true but are only used here to make a visual comparison to the CMD in Figure \ref{fig:cmd_constrained}. The position of both stars in the CMD seems consistent with the q values we have derived. In star \symbol{35}51, for example, the observed star dominates in light, hence its position in the CMD looks basically indistinguishable with that of a single star of the same mass (in agreement with isochrones). In star \symbol{35}191, the contribution of the unseen companion is more significant, but still places the binary on the turn-off of NGC 1850. 
\vspace{-0.5cm}
\section{Discussion and Conclusions}
\label{sec:concl} % used for referring to this section from elsewhere
In this paper we presented the first characterization of the binary content of NGC 1850, a 100 Myr-old and massive star cluster in the LMC. This work took advantage of a multi-epoch spectroscopic campaign performed with the IFS MUSE mounted on the VLT. Thanks to 16 epochs of observations spanning a temporal baseline of more than two years (754.1 d), we were able to identify a sample of 143 possible binaries in the MUSE FOV. This allowed us to derive an observed spectroscopic binary fraction of 10.6 $\pm$ 1.8 \% for all stars with F814W$<$19.5 in NGC 1850. This fraction was corrected for incompleteness (reflecting the limitations of the observational setup of our survey) thanks to the comparison with two sets of simulations, which differed for the distributions adopted for period (uniform in $\log P$ vs power law), mass ratio (uniform vs power law) and eccentricity (power law but with two different exponents). We obtained a detection probability of 43.4 $\pm$ 4.7 \% from the analysis of the first set of simulations, while $\rm 45.1^{\rm -15.3}_{\rm +9.2}$ \% for the second set. We estimated a true binary fraction of 24.4 $\pm$ 5 \% and $\rm 23.5^{\rm -15.4}_{\rm +9.4}$ \%, respectively, which is smaller than what derived for other clusters in the literature (e.g. over 50\% in 30 Doradus, \citealt{Sana2013,2015A&A...580A..93D} and NGC 6231, \citealt{Banyard2022}, over 40\% in Westerlund 1, \citealt{Ritchie2021} and about 34\% in NGC 330, \citealt{Bodensteiner2021}) where much more massive stars were sampled. NGC 1850 nicely fits into this picture, proposed by \citet{MoeDiStefano2017}, that more massive stars tend to show higher binary fractions.

We also investigated whether the intrinsic binary fraction of NGC 1850 changed as a function of mass and after correcting for our efficiency in detecting binaries at different magnitudes, we still observed a trend, with a decrease in the intrinsic binary fraction from 48 \% to 17 \%, for magnitudes in the range 15.5 $<$ F814W $<$ 19.5. The same behavior was also observed when splitting the NGC 1850 sample in two mass regimes. In fact, stars with intermediate-masses (2.5 $<$ M $<$ 4 $M_{\odot}$) have an intrinsic binary fraction of $f_{\rm SB, \rm corr}$ = 18.1 $\pm$ 4.2 \%, while stars with high-masses ($>$4 $M_{\odot}$) have $f_{\rm SB, \rm corr}$ = 29.5 $\pm$ 3.7 \%. When comparing the high-mass value with that found by \citet{Bodensteiner2021} in NGC 330 (34 $\pm$ 8 \%), where they sample stars down to the same mass level (M = 4 $M_{\odot}$) we get similar results, confirming that the trend in mass observed in other clusters is also manifesting in NGC 1850.

By exploiting the high flexibility of \textsc{The Joker}, a software designed to infer binary properties from sparse radial velocity curves, we were able to analyze our clean dataset of binaries and to directly constrain the orbital properties (period, eccentricity, semi-amplitude velocity K etc.) for a sample of 27 binary systems in the MUSE FOV, all of which are likely cluster members, except for two which seem to belong to NGC 1850B. By tracing the distribution of these parameters we found that the sensitivity of our survey is not the same for all orbital periods but significantly depends on the sampling of our epochs, with the intervals between 10 d and 30 d and 200 and 400 d the least sensitive. The peak at $\sim$ 400-500 d, i.e. an excess of points in this range, is almost entirely produced by the binaries constrained thanks to additional epochs from the WiFeS data. Of all the Keplerian parameters we have derived, the eccentricity is the most uncertain, although a preference for low eccentric orbits appears to be present.

Despite the limited number of epochs available, the MUSE survey has allowed the characterization (in terms of individual orbital properties) of approximately 17\% of all binaries included in the final clean dataset of NGC 1850 stars with reliable radial velocity measurements. Among the sample of binary systems we identified several interesting objects such as two eclipsing binaries, a subdwarf O-type star candidate and four objects showing a binary mass function f(M)$>$1.25 $M_{\odot}$, i.e NGC 1850 systems in which the invisible component is more massive than the observed star. One of these systems is NGC 1850 BH1, an ellipsoidal binary with a peak-to-peak radial velocity variation exceeding 300 $\,{\rm km\,s^{-1}}$ which has been extensively discussed in \citet{Saracino2022,saracino2023}. Of the three remaining systems, star \symbol{35}47 is really peculiar. It shows double-peaked emission both in the Balmer lines (H$\alpha$ and $H\beta$) and in the Paschen series and its radial velocity curve is consistent with a rather long orbital period (P = 508.4 d or P = 984.3 d). Both the very high binary mass function of the system (f(M) = 10.54 $M_{\odot}$ or f(M) = 14.46 $M_{\odot}$) and the results of the spectral disentangling indicate that this system hosts a BH candidate. Further data in different wavelength ranges will be crucial in constraining the nature of this and the other systems.

These findings confirm that we need to continue working in this direction for two main reasons. In fact, determining the orbital solutions of the detected binary systems in NGC 1850 and other clusters will allow us 1) to map the orbital parameter distributions of these stars so that the binary fractions we measure no longer need to rely on an adopted parent orbital parameter distribution as done in this work and in all other literature studies so far; 2) to investigate the presence of peculiar systems hosting dark objects such as NSs and BHs, and finally constrain their populations in clusters, so far totally unknown. The observational campaign we are conducting with MUSE will see an extension in the future, with new data and new clusters to be added to the list with the aim of shedding more light on these questions.
\vspace{-0.5cm}
\section*{Data Availability}
The data underlying this work are available in Table \ref{tab:properties} of the paper. The MUSE observations analyzed here are stored in the ESO archive, while the WAGGS images can be retrieved from the WiFeS website. The radial velocity measurements underlying this study will be shared upon reasonable request to the authors.
\vspace{-0.5cm}
\section*{Acknowledgements}
SS acknowledges funding from STFC under the grant no. R276234. SK acknowledges funding from UKRI in the form of a Future Leaders Fellowship (grant no. MR/T022868/1). MG acknowledges support from the Ministry of Science and Innovation (EUR2020-112157, PID2021-125485NB-C22, CEX2019-000918-M funded by MCIN/AEI/10.13039/501100011033) and from AGAUR (SGR-2021-01069). TS acknowledges support from the European Union's Horizon 2020 under the Marie Skłodowska-Curie grant agreement No 101024605. VHB acknowledges the support of the Natural Sciences and Engineering Research Council of Canada (NSERC) through grant RGPIN-2020-05990.

%%%%%%%%%%%%%%%%%%%%%%%%%%%%%%%%%%%%%%%%%%%%%%%%%%

\vspace{-0.5cm}
\bibliographystyle{mnras}
\bibliography{NGC1850_binaries} % if your bibtex file is called example.bib

\appendix
\section{Additional material}
\label{app:extra}
\begin{figure*}
    \centering
    \includegraphics[width=0.95\textwidth]{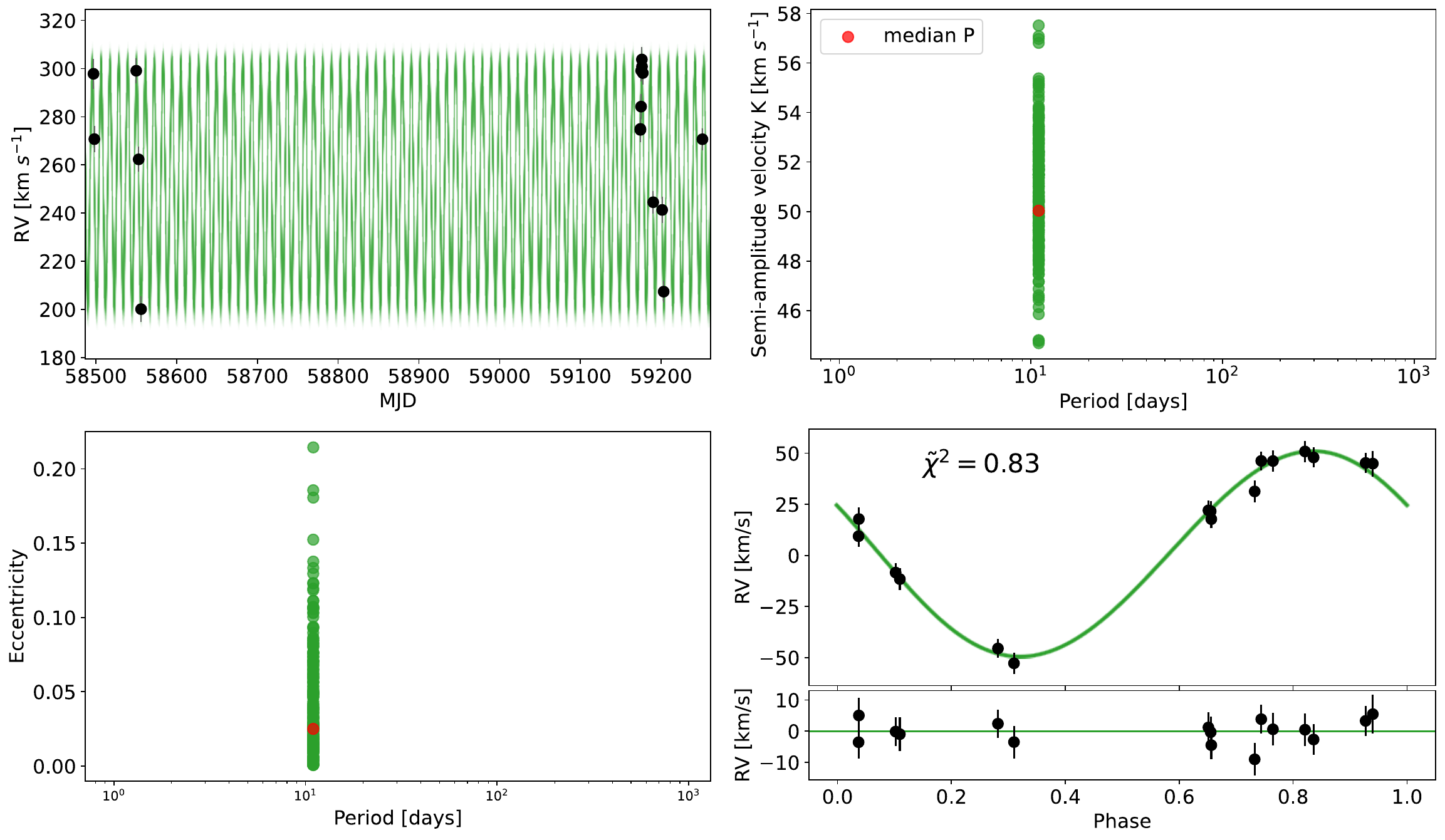}
    \caption{{\it Top-Left:} Radial velocity curve (black data points) of star \symbol{35}260 in NGC 1850. The green curves are the possible orbital solutions determined by \textsc{The Joker} for this star. {\it Top-Right:} Period - Semi-amplitude velocity K plot of these samples. {\it Bottom-Left:} Period - Eccentricity plot of the best samples. As can be seen, the orbital properties of this binary are well constrained, i.e. well clustered, this is a so-called unimodal solution. The posterior sample corresponding to the median period is shown in red, along with the values of e and K for that specific sample. {\it Bottom-Right:} Phase-folded radial velocity curve (black data points) of the binary, by assuming the best fit orbital period P = 10.97 d (see Table \ref{tab:properties}).}
    \label{fig:unimodal}
\end{figure*}

\begin{figure*}
    \centering
    \includegraphics[width=0.95\textwidth]{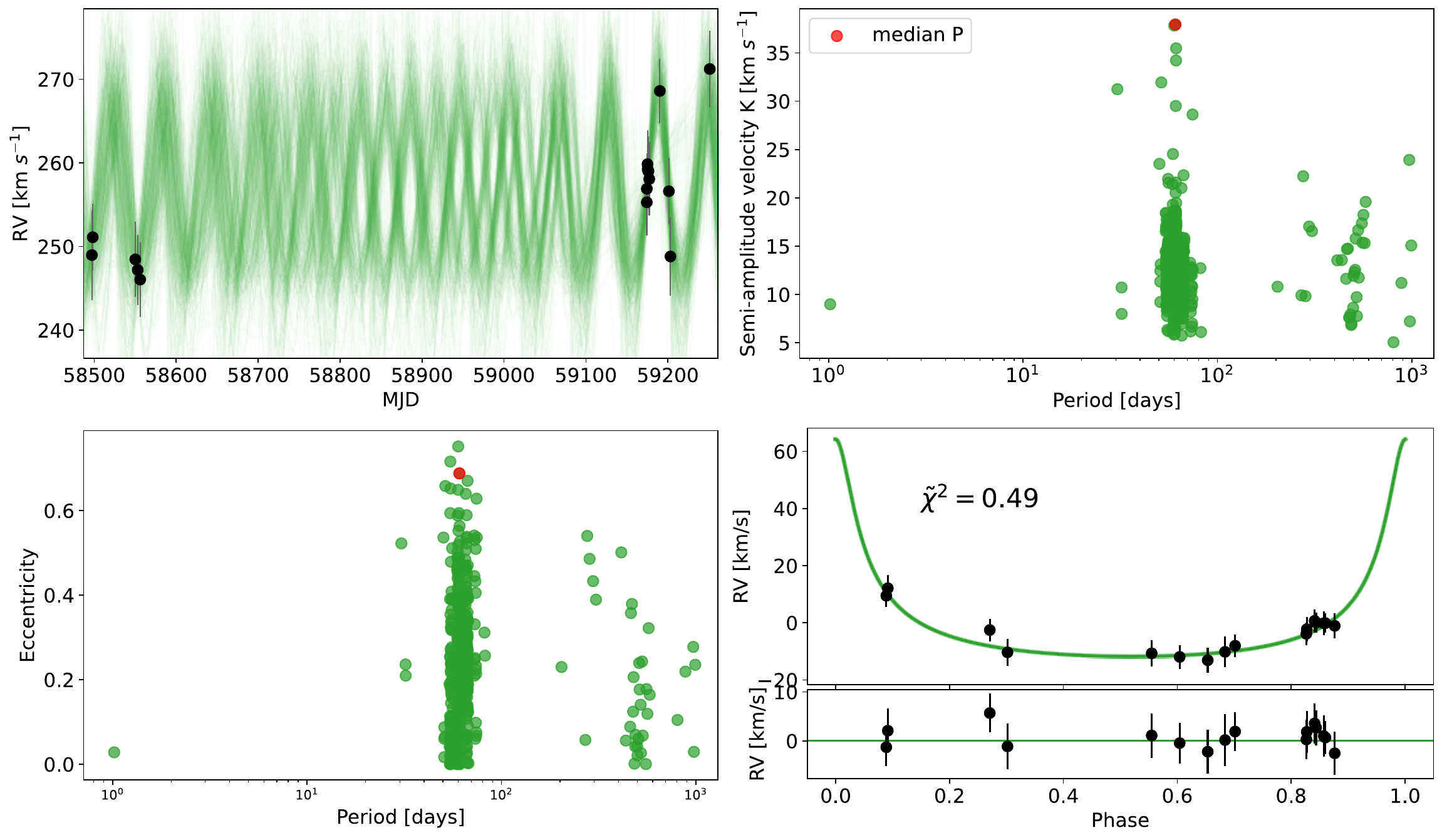}
    \caption{As in Figure \ref{fig:unimodal} but for star \symbol{35}303, that has a bimodal solution, i.e. two clusters of posterior samples well separated in the allowed period range. It is interesting to note that for the sample corresponding to the median period (red dot) has very high e and K values. It is not accurate but is only for illustrative purposes. {\it Bottom-Right:} Phase-folded radial velocity curve (black data points) of the binary, by assuming the median orbital period P = 60.48 d (see Table \ref{tab:properties}).}
    \label{fig:bimodal}
\end{figure*}

\begin{figure*}
\begin{subfigure}
    \centering
    \includegraphics[width=0.49\textwidth]{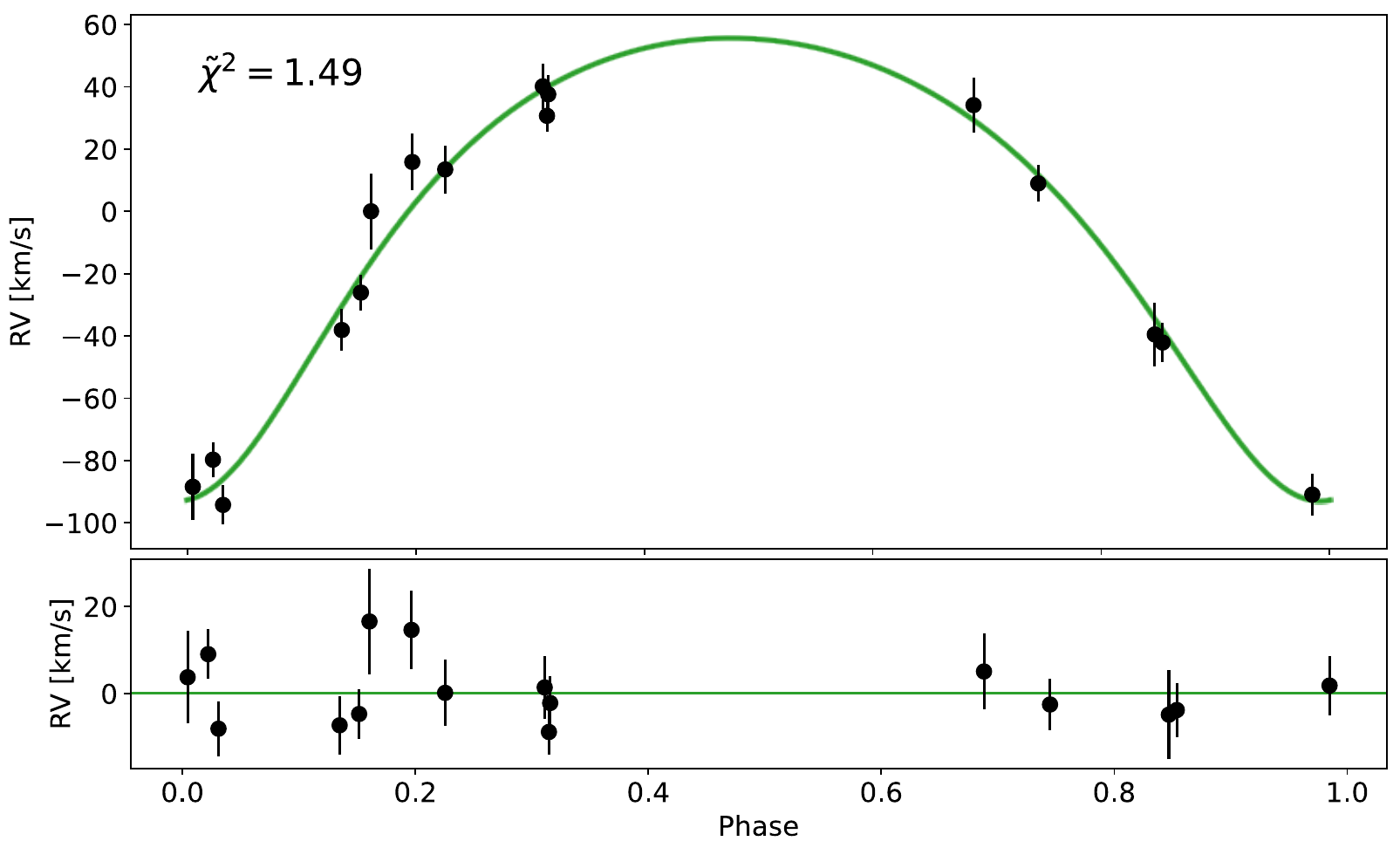}
\end{subfigure}
\begin{subfigure}
    \centering
    \includegraphics[width=0.49\textwidth]{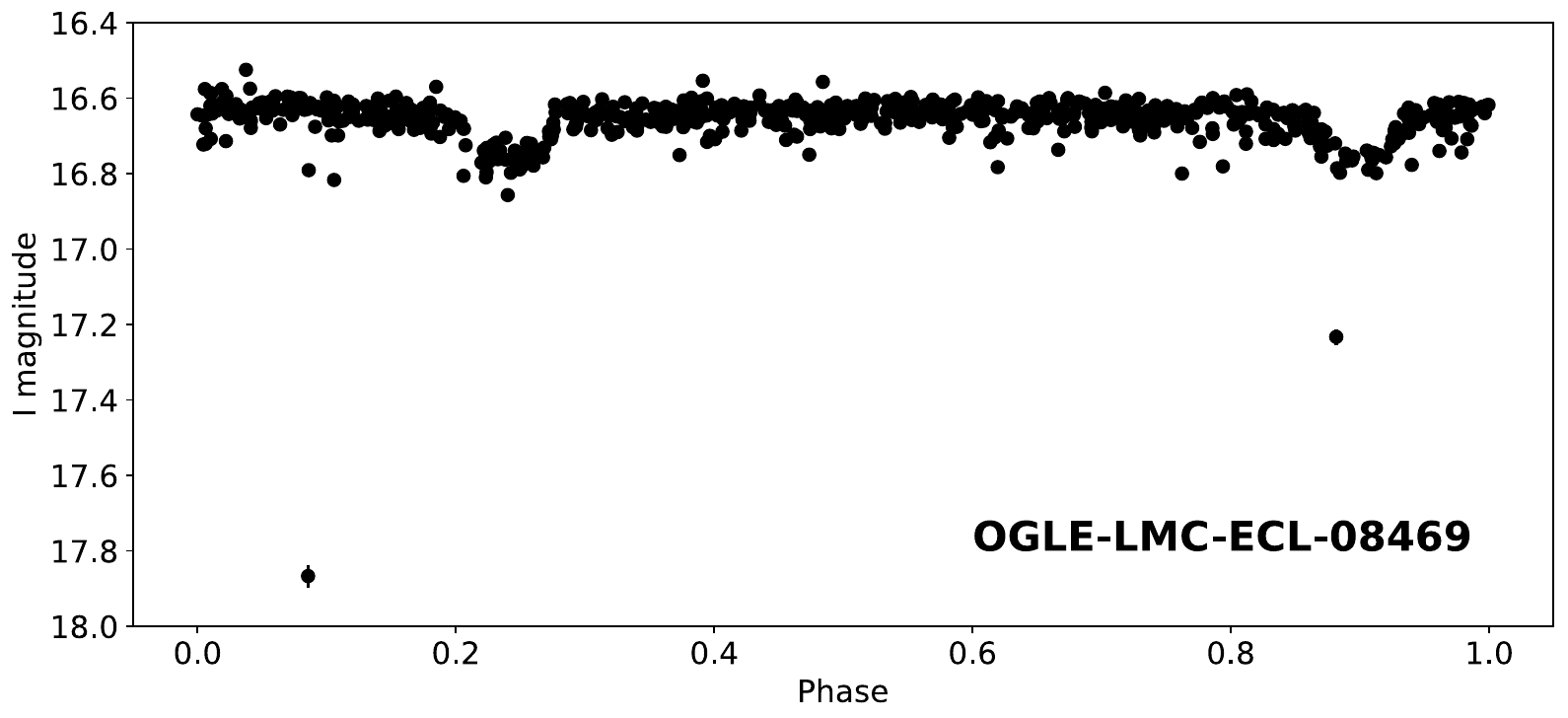}
\end{subfigure}
\caption{{Left:} Phase-folded radial velocity curve of star \symbol{35}191, as derived from MUSE radial velocity measurements, using a period P=6.46 d. This system has been also detected by OGLE-IV, classified as eclipsing binary and named OGLE-LMC-ECL-08469. {Right:} The I band light curve of the system is shown, phase-folded by using the orbital period measured spectroscopically. Two eclipses are clearly observed: a primary eclipse, which occurs when the brighter star is eclipsed by the fainter star and a secondary eclipse, which instead occurs when the fainter star is eclipsed by the brighter star.}
\label{fig:ecl}
\end{figure*}

This Appendix contains supplementary material (Table \ref{tab:properties} and Figures \ref{fig:unimodal}, \ref{fig:bimodal} and \ref{fig:ecl}) which help to better understand the content of the paper, without interrupting its flow.

\begin{table*}
  \begin{adjustbox}{addcode={\begin{minipage}{\width}}{\caption{Properties of well constrained binaries in NGC 1850.}\end{minipage}},rotate=90,center}
    \label{tab:properties}
    \begin{tabular}{p{0.5cm}p{1.2cm}p{1.45cm}p{0.6cm}p{1cm}p{1cm}p{1cm}p{0.7cm}p{0.7cm}p{0.7cm}p{0.7cm}p{0.8cm}p{0.8cm}p{0.9cm}p{0.9cm}p{0.9cm}p{0.7cm}p{0.6cm}p{0.6cm}p{0.7cm}}%{|l|l|l|l|l|l|l|l|l|l|l|l|l|l|l|l|l|l|l|l|}
    \hline \hline
        ID & RA & Dec & F438W & P\textsubscript{min} & P & P\textsubscript{max} & e\textsubscript{min} & e & e\textsubscript{max} & K\textsubscript{min} & K & K\textsubscript{max} & $V_{\rm sys,min}$ & $V_{\rm sys}$ & $V_{\rm sys,max}$ & MF\textsubscript{min} & MF & MF\textsubscript{max} & Comment \\ \hline
        23.0 & 77.1858427 & -68.7676018 & 15.77 & 108.6848 & 109.1068 & 131.6805 & 0.1105 & 0.1597 & 0.2081 & 46.0620 & 48.7677 & 53.0484 & 255.3504 & 258.7387 & 261.3333 & 1.0829 & 1.2642 & 1.9103 & unimodal \\
        27.0 & 77.1953739 & -68.7636172 & ~ & 525.3469 & 532.1627 & 539.5041 & 0.2982 & 0.354 & 0.4107 & 19.4546 & 20.0797 & 20.9085 & 259.7609 & 260.4277 & 261.1113 & 0.3493 & 0.366 & 0.3882  & unimodal \\
        32.0 & 77.1633592 & -68.7616533 & 15.94 & 7.1378 & 7.1423 & 7.2194 & 0.0944 & 0.1898 & 0.3066 & 35.5932 & 39.7537 & 44.345 & 259.6886 & 262.0882 & 264.7975 & 0.033 & 0.0441 & 0.0564  & unimodal \\
        47.0 & 77.1819232 & -68.7573088 & 15.842 & 505.6258 & 508.4290 & 511.7422 & 0.0144 & 0.0532 & 0.1255 & 54.6587 & 58.5209 & 61.7613 & -10.580 & -7.8655 & -5.1277 & 8.5719 & 10.5372 & 12.2257 & unimod,\symbol{35}1 \\
         & & & & 965.8092 & 984.3325 & 1000.79 & 0.4354 & 0.5004 & 0.5654 & 56.0669 & 60.1840 & 66.3449 & -27.965 & -24.821 & -21.511 & 12.8967 & 14.4616 & 17.0303 & unimod,\symbol{35}2 \\
        51.0 & 77.1619556 & -68.7626242 & 14.829 & 2.7779 & 3.1459 & 3.1464 & 0.2878 & 0.3809 & 0.426 & 45.0168 & 48.3666 & 51.2147 & 245.3524 & 247.8616 & 251.0311 & 0.0231 & 0.0292 & 0.0325  & unimodal \\
        100.0 & 77.1905937 & -68.7612249 & 16.12 & 4.2837 & 4.2838 & 4.284 & 0.0056 & 0.0189 & 0.0449 & 68.4437 & 69.3867 & 71.0219 & 244.7752 & 246.097 & 246.9837 & 0.1426 & 0.1485 & 0.1589 & unimodal \\
        153.0 & 77.199318 & -68.7648252 & 16.539 & 116.346 & 117.2412 & 118.3929 & 0.0461 & 0.1888 & 0.4014 & 11.9457 & 13.907 & 17.6634 & 249.1117 & 251.7891 & 254.1205 & 0.0205 & 0.031 & 0.0521 & unimodal \\
        157.0 & 77.1969601 & -68.7605334 & 15.736 & 475.2336 & 491.0016 & 507.4283 & 0.1065 & 0.1542 & 0.2128 & 28.7210 & 30.8854 & 33.4065 & 247.4456 & 250.2099 & 253.2458 & 1.1494 & 1.4490 & 1.8327 & unimodal \\
        168.0 & 77.1902562 & -68.7599008 & 17.363 & 416.4354 & 423.2189 & 428.8831 & 0.3073 & 0.3823 & 0.4466 & 8.4465 & 9.0425 & 9.739 & 255.2694 & 255.966 & 256.4861 & 0.0225 & 0.0256 & 0.0295 & unimodal \\
        191.0 & 77.2075347 & -68.7614264 & 16.65 & 6.4599 & 6.4603 & 6.4608 & 0.2563 & 0.2591 & 0.2759 & 71.3268 & 71.3732 & 72.7833 & 244.3298 & 246.7269 & 249.3449 & 0.2198 & 0.2199 & 0.2297 & unimodal \\
        210.0 & 77.1772189 & -68.7588161 & 15.61 & 439.5065 & 489.7254 & 539.019 & 0.0294 & 0.1178 & 0.2887 & 6.9675 & 8.0791 & 10.1676 & 241.0126 & 242.4883 & 244.2612 & 0.0154 & 0.0263 & 0.0516 & unimodal \\
        224.0 & 77.1945262 & -68.7654576 & 16.701 & 5.0398 & 5.0402 & 5.0406 & 0.015 & 0.029 & 0.039 & 173.0 & 175.6 & 178.2 & 250.86 & 253.30 & 255.89 & 2.71 & 2.83 & 2.97 & unimodal \\
        260.0 & 77.1846355 & -68.7614749 & 16.724 & 10.9714 & 10.9736 & 10.9762 & 0.0117 & 0.0318 & 0.0749 & 48.9269 & 50.5765 & 52.7357 & 250.9027 & 252.3807 & 253.7285 & 0.1334 & 0.1472 & 0.1658 & unimodal \\
        303.0 & 77.2190218 & -68.766609 & 16.927 & 55.3308 & 60.6082 & 66.7152 & 0.0455 & 0.2106 & 0.3775 & 9.5147 & 11.721 & 14.6992 & 254.5511 & 255.8211 & 257.2768 & 0.0049 & 0.0095 & 0.0175 & bimodal \\
        309.0 & 77.2068028 & -68.7578135 & 16.888 & 73.469 & 82.1695 & 93.5101 & 0.0206 & 0.093 & 0.2043 & 15.3944 & 17.3564 & 19.653 & 245.5757 & 248.279 & 251.0417 & 0.0278 & 0.044 & 0.0691 & unimodal \\
        368.0 & 77.1939224 & -68.776444 & 17.707 & 64.9333 & 72.1027 & 73.5741 & 0.033 & 0.1228 & 0.3117 & 20.6615 & 24.34 & 28.0206 & 242.617 & 245.5319 & 248.0731 & 0.0594 & 0.1055 & 0.1442 & unimodal \\
        383.0 & 77.2003129 & -68.7665228 & 17.012 & 81.2132 & 93.0498 & 110.3727 & 0.0204 & 0.08 & 0.1926 & 17.1684 & 18.5805 & 20.7403 & 250.681 & 253.901 & 256.7475 & 0.0427 & 0.0614 & 0.0966 & unimodal \\
        406.0 & 77.1909217 & -68.7664706 & 17.151 & 42.3642 & 45.5227 & 48.9366 & 0.0248 & 0.1184 & 0.2736 & 15.2694 & 17.1268 & 19.1126 & 247.0956 & 249.0214 & 251.4545 & 0.0156 & 0.0233 & 0.0316 & unimodal \\
        411.0 & 77.1862546 & -68.7629615 & 17.056 & 5.7597 & 7.297 & 7.5592 & 0.0225 & 0.1141 & 0.2704 & 21.1105 & 26.2882 & 30.9391 & 240.6136 & 244.3205 & 247.7001 & 0.0056 & 0.0135 & 0.0207 & unimodal \\
        427.0 & 77.1876368 & -68.7544059 & 16.988 & 442.8269 & 517.9928 & 920.4288 & 0.0371 & 0.1457 & 0.3415 & 3.9003 & 5.3574 & 9.1208 & 244.6858 & 247.3806 & 249.5994 & 0.0027 & 0.0080 & 0.0602 & unimodal \\
        467.0 & 77.1946182 & -68.7612511 & 17.198 & 48.7915 & 55.8485 & 64.9666 & 0.2078 & 0.4144 & 0.5467 & 27.0536 & 35.6647 & 52.1651 & 238.9418 & 247.2756 & 255.2327 & 0.0939 & 0.1983 & 0.5623 & unimodal \\
        497.0 & 77.211228 & -68.7822897 & 17.632 & 94.008 & 94.4513 & 145.791 & 0.0295 & 0.1012 & 0.2363 & 12.5284 & 13.8214 & 15.1623 & 255.4701 & 256.3741 & 257.3986 & 0.0192 & 0.0255 & 0.0484 & unimodal \\
        543.0 & 77.1814962 & -68.771102 & 17.269 & 10.5983 & 10.6029 & 10.6058 & 0.0269 & 0.0625 & 0.1167 & 45.062 & 47.181 & 49.2887 & 245.7724 & 247.582 & 249.1789 & 0.1006 & 0.115 & 0.1292 & unimodal \\
        815.0 & 77.2154947 & -68.7717981 & 17.818 & 102.5762 & 122.8708 & 179.5909 & 0.0196 & 0.0767 & 0.1984 & 17.3968 & 21.4377 & 26.9539 & 254.4822 & 257.7386 & 262.1063 & 0.0561 & 0.1246 & 0.3439 & unimodal \\
        1280.0 & 77.2037119 & -68.76453 & 18.063 & 3.5114 & 3.5122 & 3.5126 & 0.0718 & 0.1417 & 0.2432 & 33.5771 & 35.998 & 39.9329 & 272.7501 & 274.3785 & 275.8595 & 0.0137 & 0.0165 & 0.0212 & unimodal \\
        1828.0 & 77.2153493 & -68.767218 & 18.674 & 1.3864 & 1.3865 & 1.3865 & 0.1175 & 0.1734 & 0.2074 & 52.9644 & 57.1411 & 60.7041 & 243.547 & 246.3243 & 247.9795 & 0.021 & 0.0257 & 0.0302 & unimodal \\
        4365.0 & 77.1779216 & -68.773794 & 19.649 & 2.8724 & 2.8974 & 3.7431 & 0.0292 & 0.139 & 0.3362 & 35.5751 & 42.5284 & 51.5086 & 251.1931 & 256.322 & 261.5185 & 0.0134 & 0.0225 & 0.0444 & unimodal \\
\hline \hline
    \end{tabular}
\end{adjustbox}
\end{table*}


% Don't change these lines
\bsp	% typesetting comment
\label{lastpage}
\end{document}